\documentclass[aps,superscriptaddress,prd,
onecolumn,
floatfix, 
nofootinbib,
amsmath,amssymb,amsfonts,longbibliography]{revtex4-2}

\usepackage[dvipsnames]{xcolor}
\usepackage{float}
\usepackage{physics}
\usepackage{comment}
\usepackage[normalem]{ulem}
\usepackage{caption,subcaption}
\usepackage[pdftex]{graphicx}
\usepackage{bm,latexsym,amsmath,amssymb,amsfonts,mathrsfs}
\usepackage{empheq}
\allowdisplaybreaks[1]
\usepackage[pdftex,colorlinks=true,linkcolor=blue,citecolor=cyan,backref=page]{hyperref}

\newcommand*{\Blue}{\textcolor[rgb]{0.0,0.0,0.8}}

\newcommand{\PT}{Pöschl-Teller }
\newcommand{\schrodinger}{Schr\"odinger}
\begin{document}
\title{Sudden Decoherence by Resonant Particle Excitation \\
for Testing Gravity-Induced Entanglement}
%



%

\author{Youka Kaku}
\email{kaku.yuka.g4@s.mail.nagoya-u.ac.jp}
\affiliation{Department of Physics, Graduate School of Science, Nagoya University, Chikusa, Nagoya 464-8602, Japan}

\author{Akira Matsumura}
\email{matsumura.akira@phys.kyushu-u.ac.jp}
\affiliation{Department of Physics, Kyushu University, Fukuoka, 819-0395, Japan}

\author{Tomohiro Fujita}
\email{fujita.tomohiro@ocha.ac.jp}
\affiliation{Department of Physics, Ochanomizu University, Bunkyo, Tokyo 112-8610, Japan}
\affiliation{Research Center for the Early Universe, The University of Tokyo, Bunkyo, Tokyo 113-0033, Japan}
\affiliation{Kavli Institute for the Physics and Mathematics of the Universe (WPI),
The University of Tokyo Institutes for Advanced Study,
The University of Tokyo, Chiba 277-8583, Japan}

\date{\today}

\begin{abstract}
We propose a novel method to probe gravity-induced entanglement. We consider the gravitational interaction between a particle trapped in a shallow potential and a harmonic oscillator. The harmonic oscillator is in a quantum superposition of two frequencies and only one of these states can excite the trapped particle via resonance. Once the excited particle is detected, the quantum state of the oscillator is collapsed, 
which can be observed as 
the sudden disappearance of the superposition of oscillator frequencies.
Thus, the 
sudden decoherence, which is only triggered by particle detection, can be a smoking gun evidence of gravity-induced entanglement.
Since the probability of particle excitation increases linearly with time, the total probability is multiplied by repeating experiments. 
We will also discuss experimental implementations using optomechanics.
\end{abstract}
\maketitle

\section{Introduction}

Developing a quantum theory of gravity remains one of the most profound tasks in theoretical physics. 
A main challenge is the absence of experimental evidence to test the quantum nature of gravity.
To address this, Feynman proposed a thought experiment exploring the behavior of a gravitational field when its source is placed in a quantum superposition~\cite{Feynmann}. Building on this idea, Bose \textit{et al.} and Marletto and Vedral proposed the experimental setup to verify whether the gravitational interaction mediating between spatially superposed massive objects can produce quantum entanglement~\cite{Bose2017,Marletto2017}. Their idea, collectively reffered to as the BMV proposal, relies on a fundamental concept of quantum information theory: local operations and classical communication (LOCC) cannot produce quantum entanglement between two systems. If two massive objects interacting solely through Newtonian gravity evolve from an initially separable state into an entangled state, this would imply that their gravitational interaction cannot be described as LOCC.

To illustrate the concept of the BMV proposals, we consider the Newtonian gravitational interaction between a particle with mass $m$ and another particle with mass $M$. Let us assume that the interaction is described by the first-quantized form as
\begin{align}
\hat{H}_{\text{grav}} = -\frac{GmM}{| \hat{\bm{x}} - \hat{\bm{y}}|},
\label{eq:Hgrav1}
\end{align}
where $\hat{\bm{x}}$ and $\hat{\bm{y}}$ are the position operators of each particle. 
This interaction couples the position operators of the two systems, leading to quantum entanglement between them. In this paper, we primarily focus on this quantized gravitational interaction.
As a counterpart to the quantized gravity model, alternative models of non-quantized gravity propose that gravity-induced entanglement does not occur~\cite{Kibble1978, Kibble1980, Diosi1989, Diosi2011, Penrose1996, Penrose2014, Kafri2014, Tilloy2016, Bassi2017, Carney2023, Oppenheim2023}. In these models, gravity remains fundamentally classical, although the gravitational sources are treated quantum mechanically. As a representative example of such models, we discuss the \schrodinger-Newton gravity~\cite{Bahrami2014,Anastopoulos2014,Ruffini1969} in the latter part of this paper.

The BMV proposals have paved the way for testing whether Newtonian gravitational interaction can be classified as LOCC, thereby distinguishing between various models, such as quantum gravity and semi-classical gravity models, in the low-energy regime. However, it is important to note that these proposals are unrelated to the Hilbert space or the dynamical degrees of freedom of the gravitational field. Thus, they do not provide direct evidence for the quantization of gravity. To prove the quantization of gravitational field, experiment at higher energy scales are necessary.

While the BMV proposals offer a promising approach to exploring the quantum nature of gravity in next-generation experiments, several technical challenges must be overcome to realize them. 
One of the most significant issues is that gravity is too weak compared to other noise sources, which leads to undesired decoherence in quantum experiments~\cite{Rijavec2021}. To mitigate this issue, advancements in experimental techniques have been made to preserve quantum coherence even in mesoscopic-scale systems~\cite{Panda2024, Bild2023, Fein2019}, and precise measurements of gravitational interactions in microscopic systems have been 
demonstrated~\cite{Westphal2021, Lee2020}. Additionally, various strategies have been proposed to enhance the feasibility of detecting gravity-induced entanglement through the optimization of experimental setups~\cite{Krisnanda2020, Fujita2023, Kaku2023, Pedernales2022}. 
Refs.\cite{Krisnanda2020, Fujita2023} considered released masses and inverted oscillators for efficiently generating large gravity-induced entanglement, and 
Ref.\cite{Pedernales2022} reported the enhancement of quantized gravitational interaction in a tripartite system.

As enumerated above, there are several works on enhancing quantum gravity signal. 
However, the use of resonance effects for testing quantized gravity model, as considered in our previous work \cite{Kaku2023}, is not fully explored. 
In this paper, we investigate a setup that leverages resonant enhancement and is highly sensitive to gravity-induced entanglement.
Specifically, we consider the quantized Newtonian gravitational interaction between two particles: one is trapped in a shallow potential, and the other is a harmonic oscillator with a frequency controlled by an auxiliary qubit system. 
The harmonic oscillator is initially prepared in a superposition state of the quantum-controlled frequency.
This paper highlights three key points. First, resonance plays key role to excite the initially trapped particle through gravitational interaction. Second, we trigger the wave function collapse by measuring the particle excited from the shallow potential. As a result, we demonstrate that the interference visibility of the qubit-oscillator system suddenly vanishes if we detect the gravitationally excited particle. This sudden decoherence phenomenon does not occur under the Schrödinger-Newton gravity, and thus our method can clearly distinguish between quantum and semi-classical gravity theories. 
Third, the probability of observing sudden decoherence increases with repeated experiments, implying that it is unnecessary to maintain coherence for an exceptionally long duration in each individual experiment.
Finally, we discuss potential experimental implementations using optomechanical devices.

The rest of the paper is organized as follows.
In Section~\ref{sec:setup}, we introduce the setup and demonstrate the gravitational excitation of the particle in the shallow potential after time evolution. Section~\ref{sec:sudden_decoherence} explains the details of the sudden decoherence and its relationship to gravity-induced entanglement. In Section~\ref{sec:experimental_realization}, we discuss the experimental realization using the optomechanical device. Finally, we conclude our findings in Section~\ref{sec:conslusion}.

\section{Resonant excitation of a particle by gravity
}
\label{sec:setup}

\subsection{Setup}

\begin{figure}[htbp]
    \centering
    \includegraphics[width=0.8\linewidth]{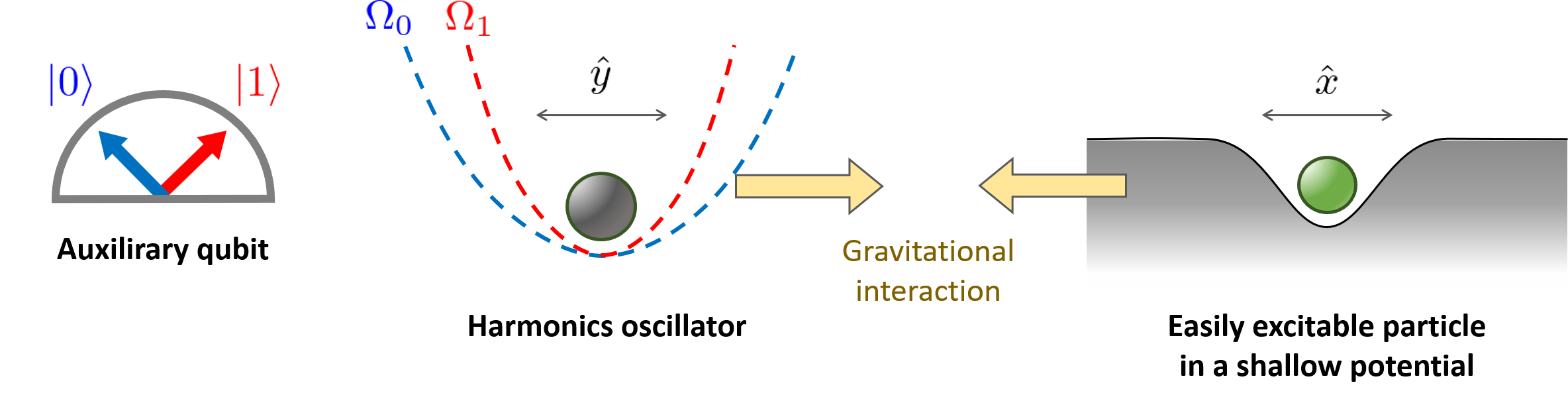}
    \caption{
    An easily excitable particle in a shallow potential (green blob on right side) interacting gravitationally with a harmonic oscillator (black blob on left side). The oscillator has two superposed 
    oscillation frequencies $\Omega_0$ and $\Omega_1$ for the control qubit states $|0\rangle$ and $|1\rangle$, respectively (leftmost arrow and harmonic potential in red and blue colors). The particle will be entangled with the oscillator system through gravitational interaction after time evolution.}
    \label{fig:setup1}
\end{figure}

We consider an excitable particle that is gravitationally interacting with a harmonic oscillator (Fig.~\ref{fig:setup1}).
The particle is initially trapped in a shallow potential.
An oscillator system is given by the superposition of two coherent states with different eigenfrequencies $\Omega_0$ and $\Omega_1$. We also introduce a qubit system state $|0\rangle,~|1\rangle$
that controls 
the oscillator frequency $\Omega_0,~\Omega_1$, respectively. 
The gravitational interaction between the particle and the oscillator systems is described by the first quantized form of the Newtonian potential, which can generate quantum entanglement between the two systems.

The Hamiltonian of the total system is given by
\begin{align}\label{eq:Hamiltonian}
    \hat H=\hat H_{\text{PT}}+\hat H_{\text{osc}}+\hat H_{\text{grav}}\,.
\end{align}
These three terms 
are the Hamiltonian of the particle in a shallow potential, 
the Hamiltonian of the oscillator, and the gravitational interaction between the particle and the oscillator systems. 
We will explain them in order.

First, we model the shallow potential of the easily excited particle by the \PT potential~\cite{Poschl1933}.
This potential allows us an analytical treatment.
Then, the Hamiltonian of the particle reads
\begin{align}\label{eq:PT_Hamiltonian}
    \hat H_{\text{PT}}
    =\frac{1}{2m}\hat p_x^2 +V(\hat x),\quad
    V(\hat x)=-\frac{\hbar^2}{2mL^2}\frac{2}{\cosh^2(\hat x/L)}\,,
\end{align}
where $\hat{x}, ~\hat p_x$ are the canonical variables of the particle, $m$ is the particle mass and $L$ 
characterizes the width of the \PT potential.
For convenience, we introduce dimensionless canonical variables, $\hat X:=\hat x/L,~\hat P_x:=\hbar^{-1}L\hat p_x$,
and rewrite the above Hailtonian as
\begin{align}\label{eq:PT_Hamiltonian2}
    \hat H_{\text{PT}}
    =\frac{\hbar^2}{2m L^2}\left(\hat 
    P_x^2-\frac{2}{\cosh^2(\hat X)}
    \right)\,.
\end{align}
It is remarkable that this system has only one bound state $|b\rangle$ whose eigenenergy and wavefunction are
\begin{align}
    E_b=-\frac{\hbar^2}{2mL^2}
    =: \hbar \omega_b,
    \quad
    \psi_b(X)=\langle X|b\rangle
    =\frac{1}{\sqrt{2}\cosh(X)}\,.
    \label{def wb}
\end{align}
In contrast, the excited (unbound) states of the particle $|k\rangle$ have a continuous spectrum,
\begin{align}
    E_k=\frac{\hbar^2}{2mL^2}k^2
    =: \hbar \omega_k
    ,\quad
    \psi_k(X)=\langle X |k\rangle
    =\frac{\tanh(X)-ik}{\sqrt{2\pi}(1-ik)}e^{ikX}\,,
    \label{def wk}
\end{align}
for $-\infty < k < \infty$.
Note that these eigenstates are orthogonal and normalized as
\begin{align}
    \langle b|b\rangle=1,\quad
    \langle b|k\rangle =0,\quad
    \langle k'|k\rangle = \delta(k-k')\,.
\end{align}
Further details on the \PT potential system can be found in Appendix~\ref{sec:PTpotential}.

Next, the Hamiltonian of the oscillator system is given by
\begin{align}
    \hat H_{\text{osc}}
    =\frac{1}{2M}\hat p_y^2+\frac{1}{2}M\hat \Omega^2 \hat y^2\,,
\end{align}
where $\hat y, \hat p_y$ are the canonical variables of the oscillator system and $M$ is the oscillator mass. 
$\hat \Omega$ is the eigenfrequency of the oscillator system 
and takes $\Omega_0$ or $\Omega_1$ depending 
on the control qubit system state $|0\rangle$ or $|1\rangle$,
\begin{align}
\hat\Omega=\Omega_0|0\rangle\langle0|+\Omega_1|1\rangle\langle1|\,.
\end{align}
Again, we introduce dimensionless variables $\hat P_y:=s_y\hat p_y/\hbar,~\hat Y:=\hat y/s_y$, where $s_y=(M\Omega_0/\hbar)^{-1/2}$. Then the Hamiltonian is rewritten as
\begin{align}
    \hat H_{\text{osc}}
    =\frac{\hbar \Omega_0}{2}\left(\hat P_y^2+\left(\frac{\hat\Omega}{\Omega_0}\right)^2\hat Y^2\right)\,.
\end{align}
In the section \ref{sec:experimental_realization}, we will discuss an example of realizing the superposition of eigenfrequencies using an optomechanical system.
In the optomechanical system, the quantum superposition of a single photon gives the quantum superposition of a radiation pressure. A mechanical oscillator feeling the radiation pressure effectively has a superposed frequency.

Finally, the gravitational interaction between the particle and the oscillator is given by
\begin{align}\label{eq:gravitational_interaction}
    \hat H_{\text{grav}}
    =-\frac{GmM}{|d+\hat x-\hat y|}
    \simeq \frac{\hbar^2 g}{2 m L^2}\hat X \hat Y\,,
\end{align}
where $d$ denotes the distance between the original positions of the two particles, and $g:=2GmML s_y/\left(|E_b|d^3\right)$ is a dimensionless gravitational coupling constant. 
In the second equality, assuming $d\gg |\hat x- \hat y|$ and using the Taylor expansion, we focus on the leading interaction term
between $\hat X$ and $\hat Y$.
The neglected terms such as $\hat X,~\hat Y,~\hat X^2,~\hat Y^2,\cdots$ do not affect our main result. 

\subsection{Time evolution under gravitational interaction}

We prepare the total system at early time $t<0$ as
\begin{align}\label{eq:initial_state}
    |\Psi(t<0)\rangle
    =|\Psi_b(t)\rangle 
    :=e^{-i\omega_b t}|b\rangle_{\rm p} \otimes
    \frac{1}{\sqrt{2}}\left(|
    0
    \rangle_{\rm q} |\alpha e^{-i\Omega_0 t}\rangle_{\rm o} +|
    1
    \rangle_{\rm q} |\alpha e^{-i\Omega_1 t}\rangle_{\rm o}\right)\,,
\end{align}
where the subscripts p,q,o denote the state of the particle, control qubit, oscillator, respectively.
Here, $\alpha$ denotes the coherent parameter of the oscillator system at $t=0$.
The control qubit and the oscillator are entangled but the particle trapped in the \PT potential is uncorrelated with them.
We introduce a condition between the eigenfrequency of the oscillator $\Omega_{0,1}$ and the bound state energy of the particle $\omega_b$,
\begin{align}\label{eq:resonance_condition}
    \Omega_0<\left|\omega_b\right|<\Omega_1\,,
\end{align}
which intuitively means that only the oscillator with the higher eigenfrequency $\Omega_1$ can excite the particle in the bound state but one with $\Omega_0$ does not provide enough energy to do so (while of course the latter excitation can occur quantum-mechanically with a suppressed probability). 

The time-evolved state at $t>0$ 
calculated from the total Hamiltonian~\eqref{eq:Hamiltonian} is given by
(derivations can be found in the Appendix~\ref{sec:apdx_oscillator})
\begin{align}\label{eq:time_evolved_state}
    |\Psi(t)\rangle
    &= |\Psi_b(t)\rangle
    + g\,\alpha\,
    |\text{ex}(\Omega_1)\rangle_{\rm p} \otimes \frac{1}{\sqrt{2}}|1\rangle_{\rm q} |\alpha e^{-i\Omega_1 t}\rangle_{\rm o}
    +g\, |\text{off-res}\rangle
    +\mathcal{O}(g^2)\,,
\end{align}
where the first term is the time evolution without the gravitational interaction, so it is same as the initial state~\eqref{eq:initial_state}, and the second and third terms denote the leading contribution of the gravitational interaction. The excited particle state $|\text{ex}(\Omega_1)\rangle$ in the second term is defined as
\begin{align}\label{eq:ex}
    |\mathrm{ex}(\Omega_1)\rangle
    :=\int_{-\infty}^{\infty} dk e^{-i\omega_k t}c_k(\Omega_1)|k\rangle\,,
\end{align}
with 
\begin{align}\label{eq:ck}
    c_k(\Omega_1):=\frac{|\omega_b|J_k}{\sqrt{2}}\,\frac{1-e^{i(\omega_k-\omega_b-\Omega_1)t}}{\omega_k-\omega_b-\Omega_1},
    \qquad
    J_k
    :=\langle k|\hat X|b\rangle
    =\frac{\sqrt{\pi}}{2}\frac{1}{1+ik}\frac{1}{\cosh(k\pi/2)}\,.
\end{align}
Here, $c_k(\Omega_1)$ is a transition amplitude from the ground state to the excited state of the particle with wavenumber $k$ (up to the constant factor of $g|\alpha|/\sqrt{2}$).
The expression of $c_k(\Omega_1)$ clearly indicates that the excitation process is resonantly enhanced for $\omega_k-\omega_b-\Omega_1=0$.
When the particle transits from the bound state of the energy $\omega_b$ to the excited one of $\omega_k (=-k^2 \omega_b)$ with the resonant wavenumber,
\begin{align}\label{eq:resonant wavenumber}
    |k|=k_\text{res}:=\sqrt{\frac{\Omega_1-|\omega_b|}{|\omega_b|}}\,,
\end{align}
the energy conservation is classically respected.
We emphasize that no parameter tuning is required to take advantage of resonant excitation in our setup, because a excited mode with $k=\pm k_{\rm res}$ is always available in its continuous spectrum as long as $\Omega_1>|\omega_b|$. 

Moreover, the factor of $J_k$ in Eq.~\eqref{eq:ck} represents the overlap of the wavefunctions of before and after the transition, and it decays exponentially for $|k|\gtrsim 1$,
because the spatially uniform force $\propto \hat{Y}$ acting on the particle 
(see Eq.~\eqref{eq:gravitational_interaction}) does not create modes with a large wavenumber.
Hence, the excitation at the resonant frequency is more likely to occur if the following condition is satisfied:
\begin{align}\label{eq:resonance_excitation_condition}
    0\leq k_\text{res} \ll 1\,.  
\end{align}
In other words, it is preferable to set $\Omega_1$ to be slightly larger than $|\omega_b|$ in order to amplify the resonance excitation.

The third term in Eq.~\eqref{eq:time_evolved_state} represents the contribution of off-resonant excitation.
Since this term is not enhanced by the resonance effect and stays sub-leading, we ignore it in the following discussion.
In Appendix~\ref{sec:apdx_oscillator}, we present the explicit expressions of the off-resonant terms and establishes a sufficient condition for neglecting them relative to the resonant contributions.

\subsection{Excitation probability of the bound particle}
\label{sec:excitation_prob}


Using the evolved state~\eqref{eq:time_evolved_state}, the transition probability of the particle from the initial bound state $|b\rangle$ to some excited state $|k\rangle$ is calculated as
\begin{align}\label{eq:excitation_probability}
    P_{\text{ex}}(t):=
    \int_{-\infty}^{\infty} dk\,\langle k|
    {\rm Tr}_{\rm q,o} \big[
    |\Psi(t)\rangle \langle \Psi(t)|\big]|k\rangle
    = g^2 |\alpha|^2 |\omega_b|^2
    \int_{-\infty}^{\infty} dk \left|J_k\right|^2
    \Delta_k(t)\,,
\end{align}
where $\Delta_k(t):=\left(\frac{\sin[(\omega_k-\omega_{k_\text{res}})t/2]}{\omega_k-\omega_{k_\text{res}}}\right)^2$ and ${\rm Tr}_{\rm q,o}[...]$ denotes the partial trace of the oscillator and the control qubit. 
In the long time limit, $\Delta_k(t)$ is approximated by the delta functions at the peaks $k=\pm k_\text{res}$,
\begin{align}\label{delta function approximation}
    \Delta_k(t\to\infty)
    = \frac{\pi t}{4 k_\text{res}|\omega_b|}\left(\delta(k-k_\text{res})+\delta(k+k_\text{res})\right)\,.
\end{align}
Then, performing the $k$ integral in Eq.~\eqref{eq:excitation_probability}, we obtain
the excitation probability in the form of Fermi's golden rule as
(see Appendix~\ref{sec:t_saturation} for derivation)
\begin{align}\label{eq:excitation_probability_linear}
    P_{\text{ex}}(t\gtrsim t_\text{sat})\simeq 
        \frac{\pi^2 g^2 |\alpha|^2}{8k_\text{res}}\, |\omega_b|t\,,
\end{align}
where we used $|J_{k_\text{res}}|^2\simeq \pi/4$ that is valid under the condition~\eqref{eq:resonance_excitation_condition}.
As we discuss in detail in the Appendix~\ref{sec:t_saturation}, 
there is a lower bound on the time to apply the approximation~\eqref{delta function approximation}, which is 
\begin{align}\label{t_sat_def}
    t_\text{sat}:=\frac{\pi}{k_\text{res}\,|\omega_b|}\,.
\end{align}
Hence, Eq.~\eqref{eq:excitation_probability_linear} is valid only if the duration of one experiment $\tau_1$ is longer than the above saturation time, $\tau_{\rm 1}\gtrsim t_{\rm sat}$.
As long as the oscillator and the particle phase can oscillate for quite a few times within the one experiment time, $|\omega_b|\tau_1\gg 2\pi$, the saturation condition is always achievable by tuning $\omega_b$ and $\Omega_1$ such that the resonant wavenumber takes 
\begin{equation}
\frac{\pi}{|\omega_b|\tau_1}\lesssim k_\text{res} \ll 1\,. 
\end{equation}
Moreover, as derived in Appendix~\ref{sec:t_saturation}, the optimal choice to maximize the total excitation probability is to set the single experiment time as 
$\tau_1\simeq t_{\rm sat}$. 
In this case, let us suppose that the experiment is repeated $N$ times. Then, the probability of observing at least one excitation over the total duration time $t_{\rm tot}=N\tau_1$ is given by

\begin{align}
    P_\text{tot}
    &:=1-(1-P_\text{ex}(\tau_1))^N
    \simeq N P_\text{ex}(\tau_1)\\
    &\simeq 100\%~\frac{m}{M}
    \left(\frac{M/d^3}{20\,\mathrm{g/cm^3}}\right)^2
    \left(\frac{|\alpha|}{0.7}\right)^2
    \left(\frac{|\omega_b|}{1\,\mathrm{Hz}}\right)^{-2}
    \left(\frac{\tau_1}{12\,{\rm hours}}\right)
    \left(\frac{t_{\rm tot}}{1\,\mathrm{year}}\right)\,,
    \label{eq:excitation_probability_value}
\end{align}
where 
we assumed $P_\text{ex}(\tau_1)\ll1$ in the first line. In the second line, we used
$\Omega_0\simeq |\omega_b|$ and $k_{\rm res}\simeq \pi/(|\omega_b|\tau_1)\approx 7\times 10^{-5}(\frac{|\omega_b|}{1{\rm Hz}})^{-1}(\frac{\tau_1}{12{\rm hours}})^{-1}$.
The detailed derivation can be found in Appendix~\ref{sec:t_saturation}.
The advantage of our proposal is that since the excitation probability $P_\text{ex}(\tau_1)$ increases linearly in time $\tau_1$, even if the excitation probability per experiment is tiny ($0.13\%$ for the above parameters), the total probability $P_\text{tot}$ can be greatly enhanced by repeating the experiment many times. For instance, by performing the experiment, which takes half a day, over about $t_{\rm tot}/\tau_1\simeq 700$ times, we can almost certainly observe the particle excitation and probe the quantum nature of gravity.


\section{Measurement scheme and implications}
\label{sec:sudden_decoherence}

In this section, we discuss the measurements that we make in our setup and the implications for the quantum nature of gravity that we can draw from the results. 
If the Newtonian gravity is quantized, one should observe that the 
quantum superposition
of the oscillator system suddenly vanishes when the particle excited by the resonant gravitational interaction is detected. However, if the gravity is semi-classical, such a radical change of the quantum state of the oscillator system due to the particle detection is absent, as we will explicitly show below.

\subsection{Sudden decoherence 
of 
the qubit-oscillator system
}
\label{sec:radical_decoherence}

Let us consider the interference between the superposed state of the control qubit, $|0\rangle_{\rm q}$ and $|1\rangle_{\rm q}$.
The degree of interference is quantified by the interference visibility.
Calculating the interference visibility
for the time-evolved state~\eqref{eq:time_evolved_state}, we find
\begin{equation}\label{eq:V_oscillate}
    \mathcal{V}(t)
    :=2\left|\mathrm{Tr}_{\rm po}\big[{}_{\rm q}\langle1|\Psi(t)\rangle\langle \Psi(t)|0\rangle_{\rm q}\big]\right|\\
    \simeq
        \exp\left[-2\alpha^2\sin^2\left(\frac{\Omega_1-\Omega_0}{2}t\right)\right]\,.
\end{equation}
It oscillates with a maximum value of one, simply due to the difference in the superposed frequencies of the oscillator. Since the particle state is orthogonal, $\langle b|k\rangle =0$, the contribution from the resonant gravitational excitation state (i.e. the second term in Eq.~\eqref{eq:time_evolved_state}) drops.

\begin{figure}[htbp]
    \begin{tabular}{c}
      \begin{minipage}[t]{0.95\linewidth}
        \centering
        \includegraphics[width=0.8\linewidth]{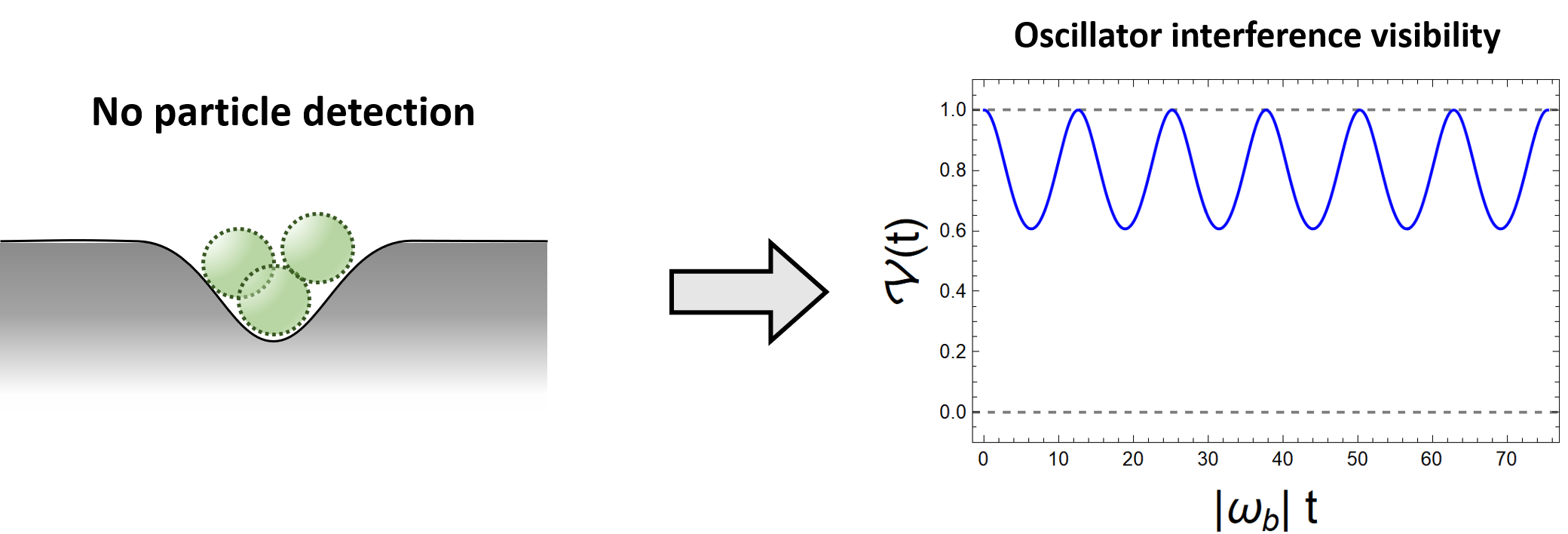}
        \subcaption{While we do not detect the excited particle, the oscillator interference visibility 
        oscillates as we see in Eq.~\eqref{eq:V_oscillate}. Here, we set $\alpha=0.5,~\Omega_1-\Omega_0=0.5|\omega_b|$.}
      \end{minipage}\\
   
      \begin{minipage}[t]{0.95\linewidth}
        \centering
        \includegraphics[width=0.8\linewidth]{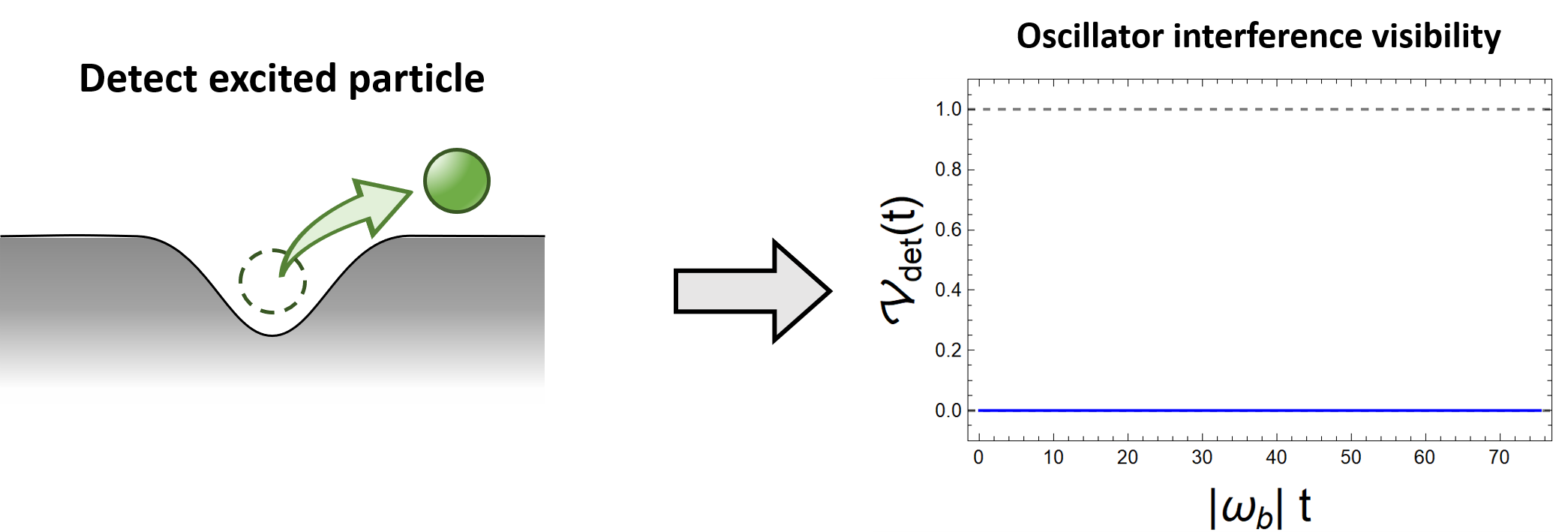}
        \subcaption{When we detect the excited particle, the interference visibility suddenly vanishes 
        as shown in Eq.~\eqref{eq:V_vanish} 
        }
      \end{minipage}
    \end{tabular}
     \caption{
     Particle detection triggers the sudden disappearance of interference visibility}
     \label{fig:radical_decoherence}
  \end{figure}

Now we consider the detection of the excited particle. We place a particle detector away from the origin of the \PT potential, which detects excited particle as it reaches that position. Provided the detected particle state is $|{\rm det}\rangle_{\rm p}$, it is orthogonal to the bound state $|b\rangle_{\rm p}$ but has an overlap with the excited state $|\rm{ex}(\Omega_1)\rangle_{\rm p}$.
Then by the projection measurement, the state immediately after the detection contains only the descendant of the particle excited state and it reads
\begin{equation}
    |\Psi_{\rm det}\rangle = \mathcal{N} |\rm{det}\rangle_{\rm p} |1\rangle_{\rm q} |\alpha e^{-i\Omega_1 t}\rangle_{\rm o}\,,
\end{equation}
where $\mathcal{N}$ is an appropriate normalization factor.
Therefore, right after the particle detection, the interference visibility disappears
\begin{equation}\label{eq:V_vanish}
    \mathcal{V}_{\rm det}
    = 0\,.
\end{equation}
In Fig.~\ref{fig:radical_decoherence}, we illustrate this sudden disappearance of the interference visibility which is triggered by the detection of the excited particle.

It should be stressed that the particle and the oscillator system are coupled only through gravitational interaction in our setup. As we have just seen, however, the measurement on the particle has significantly changed the quantum state of the oscillator system.  This implies that quantum correlations between them have been generated by gravity.
Moreover, in principle, this phenomenon occurs instantaneously even if the position
of particle measurement and the oscillator system are separated by a large distance. Thus, this successive occurrence of particle detection and loss of interference visibility will be smoking-gun evidence that gravity produces quantum entanglement between them.

\subsection{Gravity-induced entanglement}\label{sec:GIE}

The measurement procedure discussed above implied that gravity generated the entanglement between the particle and the other parts of the system. Here, we compute the negativity~\cite{Zyczkowski1998,Eisert2003} between them to quantify the gravity-induced entanglement.

Neglecting the off-resonance terms and $\mathcal{O}(g^2)$ terms, the time evolved state~\eqref{eq:time_evolved_state} is rewritten as
\begin{align}\label{eq:time_evolved_state_approx}
    |\Psi(t)\rangle
    &\simeq \frac{1}{\sqrt{2}}\Big[
    e^{-i \omega_b t}|b\rangle_{\rm p}\otimes \left(|e_0\rangle_{\rm qo}+|e_1\rangle_{\rm qo}\right) 
    + g\alpha |\mathrm{ex}(\Omega_1)\rangle_{\rm p} \otimes |e_1\rangle_{\rm qo}
    \Big]\,,
\end{align}
where we introduced a joint orthonormal basis of the oscillator and the control qubit, $|e_0\rangle_{\rm qo}:=|0\rangle_{\rm q}|\alpha e^{-i\Omega_0 t}\rangle_{\rm o}$ and $|e_1\rangle_{\rm qo}:=|1\rangle_{\rm q}|\alpha e^{-i\Omega_1 t}\rangle_{\rm o}$, respectively, which satisfy ${}_{\rm qo}\langle e_i|e_j\rangle_{\rm qo}=\delta_{ij}$.
As the above equation indicates, the entire system can be viewed as consisting of two subsystems instead of three, the particle and the other. Furthermore, we can treat them as a two-level system, with the former being the bound $|b\rangle$ or excited states $|{\rm ex}\rangle$ and the latter being $|e_0\rangle$ or $|e_1\rangle$. 

To compute the negativity between the particle and the other systems
, we first calculate the partially transposed density matrix with respect to 
the state of particle 
(or the state of the others)
using Eq.~\eqref{eq:time_evolved_state_approx}. Then, we compute the eigenvalues of the partially transposed density matrix. Finally, the negativity is given by an absolute value of a summation of all negative eigenvalues. 
Interestingly, in our setup, the negativity $N$ between the particle and the other systems coincides with the square root of the excitation probability of the particle:
\begin{align}
  N(t)=\frac{g|\alpha|}{2}\sqrt{\int_{-\infty}^\infty dk \left|c_k(\Omega_1)\right|^2}
    =\sqrt{\frac{P_\text{ex}(t)}{2}}\,.
\end{align}
Thus, when our setup is described by the quantized Newtonian gravity, measuring the particle excitation and the sudden disappearance of the visibility is nothing less than a direct observation of gravity-induced entanglement.

\subsection{The \schrodinger-Newton gravity}

Here we study the case of \schrodinger-Newton gravity as a representative example of the non-quantized gravity models.
In this model, the gravitational interaction does not create quantum entanglement between the particle and the oscillator system. We will see that although the particle can still be excited by gravitational interaction, 
the oscillator interference visibility does not vanish after the particle detection in the \schrodinger-Newton gravity case.


The \schrodinger-Newton equation for the particle is given by
\begin{align}
    i\hbar\frac{\partial}{\partial t}|\chi(t)\rangle_{\text{p}}
    &=\left(\hat H_{\text{PT}} 
    -{}_{\rm qo}\langle\varphi(t)|\frac{GmM}{\left|d+\hat x-\hat y\right|}|\varphi(t)\rangle_{\text{qo}}
    \right)|\chi(t)\rangle_{\text{p}}\,,\\
    &\simeq \left(\hat H_{\text{PT}} 
    + \hbar g\,|\omega_b|\,
    {}_{\rm qo}\langle\varphi(t)|\hat Y|\varphi(t)\rangle_{\text{qo}} \,  \hat X
    \right)|\chi(t)\rangle_{\text{p}}\,,
    \label{eq:SN_timeevolution_PT}
\end{align}
where we called the state of the particle and the qubit-oscillator system
$|\chi(t)\rangle_{\text{p}}$ and $|\varphi(t)\rangle_{\text{qo}}$, respectively and we ignored the self-gravity effect.
In the second line, we have performed the Taylor expansion for $d\gg L,~s_y$ in the same way as Eq.~\eqref{eq:gravitational_interaction}.
Since this equation is non-linear and hard to obtain its general solution, let us focus on the leading order solution in the gravitational coupling constant $g$.
Using the initial state~\eqref{eq:initial_state}, the position expectation values of 
the oscillator is ${}_{\rm qo}\langle\varphi(t)|\hat Y|\varphi(t)\rangle_{\text{qo}}
    \simeq \mathrm{Re}\left[\alpha \left(e^{-i\Omega_0 t} + e^{-i\Omega_1 t}\right)\right]/\sqrt{2}$.
Solving the above equation, we find the time-evolved state of the total system as
\begin{align}
    |\Psi^{\text{SN}}(t)\rangle
    &=|\chi(t)\rangle_{\text{p}} \otimes |\varphi(t)\rangle_{\text{qo}}\,,\\
    &=\left(
    e^{-i\omega_b t}|b\rangle_{\rm p}
    + \frac{g\,\alpha}{2} |\text{ex}(\Omega_1)\rangle_{\rm p}
    \right)
    \otimes 
    \frac{1}{\sqrt{2}}\left(
    |0\rangle_{\rm q} |\alpha e^{-i \Omega_0 t}\rangle_{\rm o}
    +|1\rangle_{\rm q} |\alpha e^{-i \Omega_1 t}\rangle_{\rm o}
    \right)
    +\text{(off-resonance)}+\mathcal{O}(g^2)\,.
    \label{eq:SN_time_evolved_state}
\end{align}
This result should be compared to the quantized gravity case in Eq.~\eqref{eq:time_evolved_state}.
In the quantized case, 
the resonance excitation of the particle occurs only when it couples to the high frequency oscillator $|1\rangle_\text{q} |\alpha e^{-i\Omega_1 t}\rangle_\text{o}$. 
However, in the \schrodinger-Newton gravity case, 
the resonance excitation occurs regardless of the oscillator states, 
and the total state remains separable.  
This is because the particle feels the averaged gravitational force of the oscillator system, and does not change its time evolution depending on the oscillator state. 
The excitation probability of the particle is obtained as
\begin{align}
    P_{\text{ex}}^{\text{SN}}(t)
    &:=\int_{-\infty}^{\infty} dk \left|\langle k|\Psi^{\text{SN}}(t)\rangle\right|^2
    \simeq \frac{1}{2}P_{\text{ex}}(t)\,,
\end{align}
which is half of the quantized Newtonian gravity case. Aside from the factor of two, the gravity-induced resonant excitation of the trapped particle occurs both in the quantized Newtonian gravity and the \schrodinger-Newton gravity cases. 

In a similar way to the section \ref{sec:radical_decoherence}, 
we compute the interference visibility before and after the detection of the excited particle in the \schrodinger-Newton gravity model as
\begin{align}
    \mathcal{V}^{\text{SN}}(t)=\mathcal{V}_{\rm dec}^{\text{SN}}(t)
    &\simeq \exp\left[-2\alpha^2\sin^2\left(\frac{\Omega_1-\Omega_0}{2}t\right)\right]\,.
    \label{eq:SN_visibility}
\end{align} 
The visibility continues to oscillate regardless of whether the excited particle is detected or not.
There is no visibility disappearance triggered by the excited particle detection in the \schrodinger-Newton gravity case, in sharp contrast to the quantized Newtonian gravity case.
Therefore, through the experiment we propose, we can determine whether gravity produces quantum entanglement, and in turn, we can distinguish the correct theory of quantum gravity in the Newtonian regime.



\section{Example of an experimental realization 
}\label{sec:experimental_realization}

\begin{figure}[htbp]
    \centering
    \includegraphics[width=0.9\linewidth]{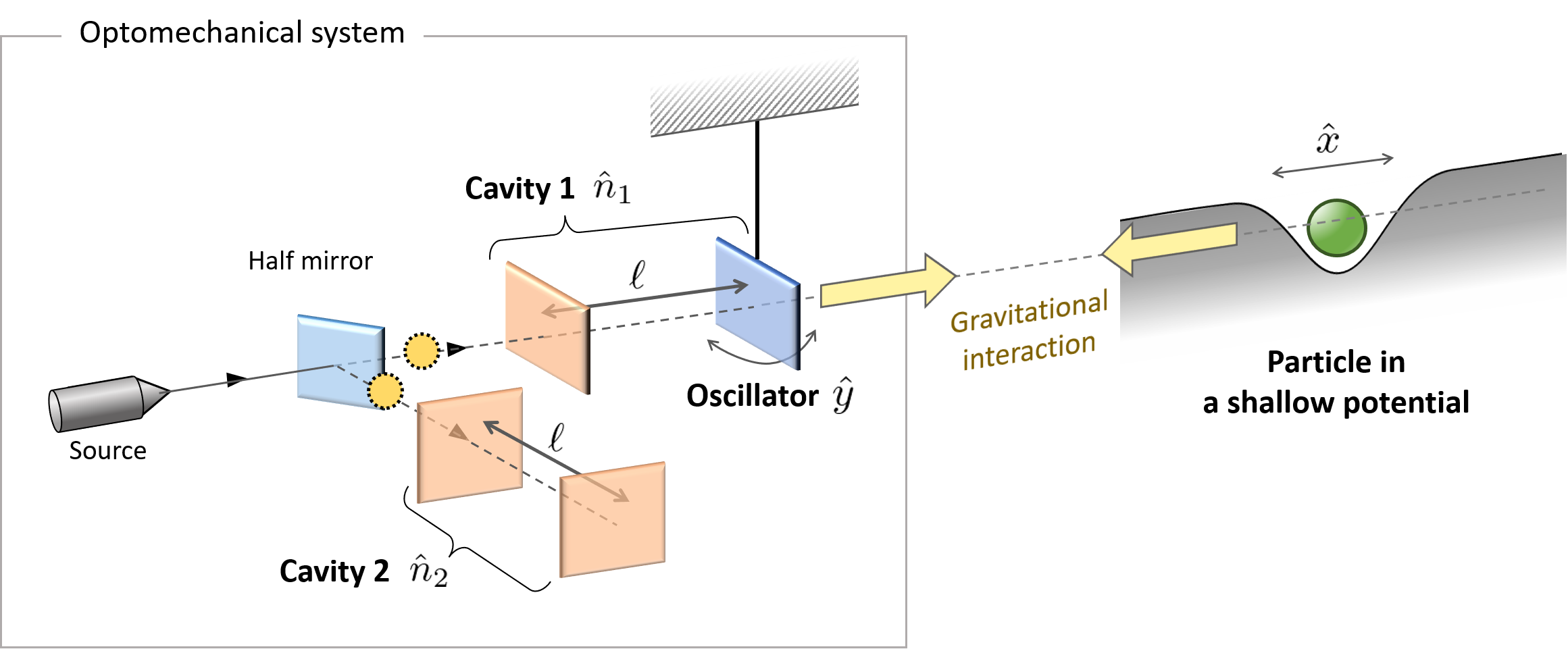}
    \caption{
    An implementation of our setup using optomechanics. A single photon enters cavities 1 and 2 as a quantum superposed states through a half mirror, which corresponds to the control qubit system introduced in the section \ref{sec:setup} (see Fig.~\ref{fig:setup1}). Only in the former case, the eigenfrequency of the oscillator is modified by the radiation pressure. The oscillator, which is thereby in a superposition of two frequencies, gravitationally couples to a particle trapped in a shallow potential.
    }
    \label{fig:setup2}
\end{figure}

In this section, we present an example of the implementation of our setup using an optomechanical cavity system. 
Fig.~\ref{fig:setup2} illustrates a schematic picture of an experimental realization. We consider an optomechanical system containing a single photon and a mechanical oscillator, where the oscillator is gravitationally interacting with an easily excitable particle. A single photon emitted from the light source splits into two directions by a half mirror and becomes a superposition of two states; the state standing in cavity 1 and cavity 2, which we denote $|1\rangle$ and $|0\rangle$ respectively.

Following our previous paper~\cite{Kaku2023}, we will briefly explain that this photon system corresponds to the control qubit system introduced in the section \ref{sec:setup}, and the oscillator changes its frequency as $\Omega_0,~\Omega_1$ depending on the photon state $|0\rangle,~|1\rangle$.
When the oscillator is in the original position $y=0$, the photon frequency inside cavity 1 is given by
\begin{align}
    \omega_c=\frac{\pi c\, \mathfrak{n}}{\ell}\,,
\end{align}
where $\ell$ is the original cavity length and $\mathfrak{n}$ is an integer. 
If the photon 
enters cavity 1, 
the mechanical oscillator 
is pushed by the photon radiation pressure and the oscillator is shifted by $y$. 
Then the photon frequency is modified as 
\begin{align}
    \omega_c'=\frac{\pi c\, \mathfrak{n}}{\ell+y}
    \simeq \omega_c\left(1-\frac{1}{\ell}y+\frac{1}{\ell^2}y^2\right),
    \label{eq:photon_frequency_modified}
\end{align}
where we performed the Taylor expansion for $y\ll \ell$ in the second equality.
Using the above form, the Hamiltonian of the photon and the oscillator systems is given by
\begin{align}
    \hat H 
    &= \hbar \omega_c' \hat n_1 + \hbar \omega_c \hat n_2
    + \frac{1}{2M}\hat p_y^2+\frac{1}{2}M \Omega_0^2 \hat y^2\,,\notag\\
    &\simeq \hbar \omega_c \left(\hat n_1 + \hat n_2\right)
    + \frac{1}{2M}\hat p_y^2+\frac{1}{2}M \hat \Omega^2 \hat y^2 - \frac{\hbar \omega_c}{\ell} \hat n_1 \hat y\,
    \label{eq:Hamiltonian_opm},
\end{align}
with 
\begin{align}
    \hat \Omega=\Omega_0|0\rangle\langle 0|+\Omega_1|1\rangle\langle 1|,\qquad
    \Omega_1 = \sqrt{\Omega_0^2+\frac{\hbar\omega_c}{M \ell^2}}\,,
\end{align}
where $\hat n_1,~\hat n_2$ is the number operator of the photon inside the cavity 1 and 2 respectively, $M$ is the oscillator mass, and $\Omega_0$ is the original frequency of the oscillator. 
We can see that the oscillator system has two different eigenfrequencies $\Omega_0$ and $\Omega_1$ depending on the photon state $|0\rangle$ and $|1\rangle$ as expected. After an optomechanical interaction has continued for a sufficiently long time, we can show that the joint state of the photon and the oscillator becomes
\begin{align}\label{eq:initial_state_opm}
    |\psi(t)\rangle_{\text{opm}}
    \simeq \frac{1}{\sqrt{2}}\left(    |0\rangle\,|\alpha_0(t)\rangle+|1\rangle\,|\alpha_1(t)\rangle\right)\,,
\end{align}
with $\alpha_0(t) \simeq e^{-i\Omega_0 t}\alpha$,
and $\alpha_1(t) \simeq e^{-i\Omega_1 t}(\alpha-\lambda_0) + \lambda_0$.
Here, $\alpha$ is an initial coherent parameter which can be an arbitrary complex number and $\lambda_0=\omega_c s_y/(2\Omega_0\ell)$ is an optomechanical coupling constant with $s_y=(M\Omega_0/\hbar)^{-1/2}$. Derivations and further details are explained in Appendix~\ref{sec:apdx_optomecha}.
Now it is evident that the photon system plays a role of the control qubit system introduced in the previous discussions. The only difference 
is that the oscillator state accompanied by the photon state $|1\rangle$ is 
shifted by $\lambda_0$ due to the photon radiation pressure. However, this difference does not affect our main result of the sudden disappearance of the interference visibility, as we will see below.

Now, let us consider the time evolution of the total system under gravitational interaction given in Eq.~\eqref{eq:gravitational_interaction}. 
We suppose that the total system at early time $t<0$ is given by
\begin{align}
    |\Psi(t<0)\rangle
    =|b\rangle\otimes|\psi(t)\rangle_{\text{opm}}
    =|b\rangle\otimes
    \frac{1}{\sqrt{2}}\left(|0\rangle |\alpha_0(t)\rangle +|1\rangle |\alpha_1(t)\rangle\right)
\end{align}
Then, by following the same argument as in the section \ref{sec:setup}, the time-evolved state is obtained as
\begin{align}\label{eq:time_evolved_state_optomecha}
    |\Psi(t)\rangle
    &= \frac{e^{-i\omega_b t}}{\sqrt{2}}|b\rangle\otimes
    \left(|0\rangle |\alpha_0(t)\rangle +|1\rangle |\alpha_1(t)\rangle\right)
    + \frac{g}{\sqrt{2}}\,\left(\alpha-\lambda_0\right)\,
    |\text{ex}(\Omega_1)\rangle \otimes |1\rangle |\alpha_1(t)\rangle
    +\left(\text{off-resonance}\right)
    +\mathcal{O}(g^2)\,,
\end{align}
where the explicit form of the omitted off-resonant terms and further details are given in Appendix~\ref{sec:apdx_optomecha}.
The difference from Eq.~\eqref{eq:time_evolved_state} is twofold; the coefficient of the second term has changed from $\alpha$ to $\alpha-\lambda_0$, and the state of the high-frequency oscillator  has changed from $|\alpha e^{-i\Omega_1 t}\rangle$ to $|\alpha_1(t)\rangle \simeq |e^{-i\Omega_1 t}(\alpha-\lambda_0) + \lambda_0\rangle$.
According to these replacements, the expressions for particle excitation probability $P_{\rm ex}$ and visibility before particle detection $\mathcal{V}(t)$ change slightly, but they do not hardly affect our main conclusions.

Finally, we comment on some practical considerations for experimental realization. In Eq.~\eqref{eq:excitation_probability_value}, we assumed a single experiment time of $\tau_1 \simeq 12\,{\rm hours}$. However, maintaining experimental coherence over such long timescales is technically challenging. For instance, a mechanical oscillator in a macroscopic superposition state easily decoheres due to air molecule collisions, with a typical timescale of $\mathcal{O}(1)\,{\rm s}$~\cite{Kaku2023}. In addition, when using an optomechanical system, a photon inside the optical cavity decays rapidly, with a characteristic timescale of
$
\tau_c = Q/\omega_c \approx 3\times 10^{-10}\,{\rm s}
\left(\frac{Q}{10^6}\right)\left(\frac{\lambda_c}{637\,{\rm nm}}\right)
$~\cite{Aspelmeyer2014,Janitz2015}.
These issues make it difficult to extend $\tau_1$ arbitrarily. Nevertheless, as discussed below Eq.~\eqref{eq:excitation_probability_value}, our proposal does not rely on a long duration of a single experiment. Instead, the key idea is to accumulate sufficient total time $t_{\rm tot}$ through multiple iterations, making the scheme experimentally more feasible.

Additionally, in the optomechanical setups, the frequency gap between $\Omega_0$ and $\Omega_1$ is determined by the single-photon radiation pressure and therefore remains very small. Following the estimation in \cite{Kaku2023}, this gap is given by
\begin{align}
    1-\frac{\Omega_0}{\Omega_1}
    \simeq \frac{\hbar\omega_c}{2M\Omega_0^2\ell^2}
    =2.8\times 10^{-10}
    \left(\frac{M}{10^{-13}\,{\rm kg}}\right)^{-1}
    \left(\frac{\Omega_0}{3\times 10^3\,{\rm Hz}}\right)^{-2}
    \left(\frac{\omega_c}{450\times 10^{12}\,{\rm Hz}}\right)
    \left(\frac{\ell}{0.01\,{\rm m}}\right)^{-2}
    ~.
\end{align}
Thus, to realize the resonant excitation under the condition $\Omega_0<|\omega_b|<\Omega_1$, the particle frequency $|\omega_b|$ must be controlled with extremely high precision, corresponding to a quality factor of about $10^{10}$, which is highly challenging.
Moreover, in order to see the sudden decoherence, the off-resonant terms in Eq.~\eqref{eq:time_evolved_state} must be properly neglected. This requirement leads to the condition
\begin{align}
    1-\frac{|\omega_b|}{\Omega_1}\ll \left(1-\frac{\Omega_0}{\Omega_1}\right)^2
    \simeq 7.8\times 10^{-20}
    ~.
\end{align}
The derivation of this condition is given in Appendix~\ref{sec:apdx_oscillator}. This indicates that $|\omega_b|$ has to be tuned extremely close to $\Omega_1$ with the quality factor of about $10^{19}$, which is far beyond what is feasible in realistic experiments. 
While we have proposed the optomechanical setup as an illustrative example to generate a superposition of different frequency states, more practical alternatives should be explored for experimental realization.


\section{Summary}\label{sec:conslusion}

Experimental evidence of the quantum gravity is highly anticipated. As a step toward investigating quantum gravity, Bose \textit{et al.} and Marletto and Vedral proposed an experimental setup to test whether gravity can create quantum entanglement between two masses~\cite{Bose2017, Marletto2017}. 
In this paper, we explored the gravitational excitation of a trapped particle via resonance, and its associated sudden decoherence, a phenomenon that is highly sensitive to gravity-induced entanglement.

Our setup, illustrated in Fig.~\ref{fig:setup1}, involves the quantized Newtonian gravitational interaction between two massive bodies: a particle trapped in a shallow potential and a harmonic oscillator 
whose frequency controlled by an auxiliary qubit system. By solving the time evolution of the total system, we demonstrated that the initially trapped particle is excited by the gravitational interaction via resonance
with a probability increasing linearly in time. 
Moreover, the measurement of the excited particle induces a corresponding state collapse.
As a result, we found that when the gravitationally excited particle is detected, the interference visibility of the qubit-oscillator system suddenly disappears. Notably, this sudden decoherence is absent in the \schrodinger-Newton gravity model, which does not produce gravity-induced entanglement. Thus, the sudden decoherence phenomenon provides a crucial indicator for testing the existence of the gravity-induced entanglement. Its feasibility is supported by the particle excitation probability, which can be accumulated through repeated experiments, as discussed in Sec.~\ref{sec:excitation_prob}.

Our framework serves as a novel method for explicitly probing gravity-induced entanglement.
In Sec.~\ref{sec:experimental_realization}, we illustrated how this framework could be implemented using an optomechanical system. However, several technical challenges remain for practical realization: First, sudden decoherence can also arise from quantum entanglement between the particle and qubit systems, mediated by an interaction other than gravity. 
Therefore, an additional treatment is required to ensure that the observed decoherence specifically originates from the gravity-induced entanglement, an issue also noted in the original BMV proposals~\cite{Bose2017, Marletto2017}.
Second, there are some experimental hurdles, such as realizing a shallow potential and reliably detecting the excitation of the trapped particle in laboratory settings. 
Overcoming these technical obstacles is crucial to fully realizing the potential of our approach. Once resolved, this framework could serve as a powerful method for testing gravity-induced entanglement.

\section*{Acknowledgments}

This work was supported in part by the Japan Society for the Promotion of Science (JSPS) KAKENHI, Grants No. JP23K03424 (T.F.), JP23K13103 (A.M.), 24KJ1233 (Y.K.) 
and by Grant-in-Aid for JSPS Fellows (Y.K.).

\appendix

\section{Eigensystem of \PT potential}
\label{sec:PTpotential}


This section will review the eigensystem of the \PT potential particle. The Hamiltonian of the particle in the \PT potential was given in Eq.~\eqref{eq:PT_Hamiltonian},
which is rewritten as Eq.~\eqref{eq:PT_Hamiltonian2}
with the dimensionless canonical variables, $\hat X:=\hat x/L$ and $\hat P_x:=\hbar^{-1}L \hat p_x$.
Let us consider the eigenstate of this Hamiltonian. If the eigenstate $|\psi\rangle$ has the eigenenergy $E$, its wave function satisfies the following \schrodinger 
~equation.
\begin{align}
    \frac{\hbar^2}{2mL^2}
    \left (\frac{d^2}{d X^2}+\frac{2}{\cosh^2(X)} \right )\psi(X)
    =E\,\psi(X)\,.
\end{align}
\begin{figure}[htbp]
    \centering
    \includegraphics[width=0.7\linewidth]{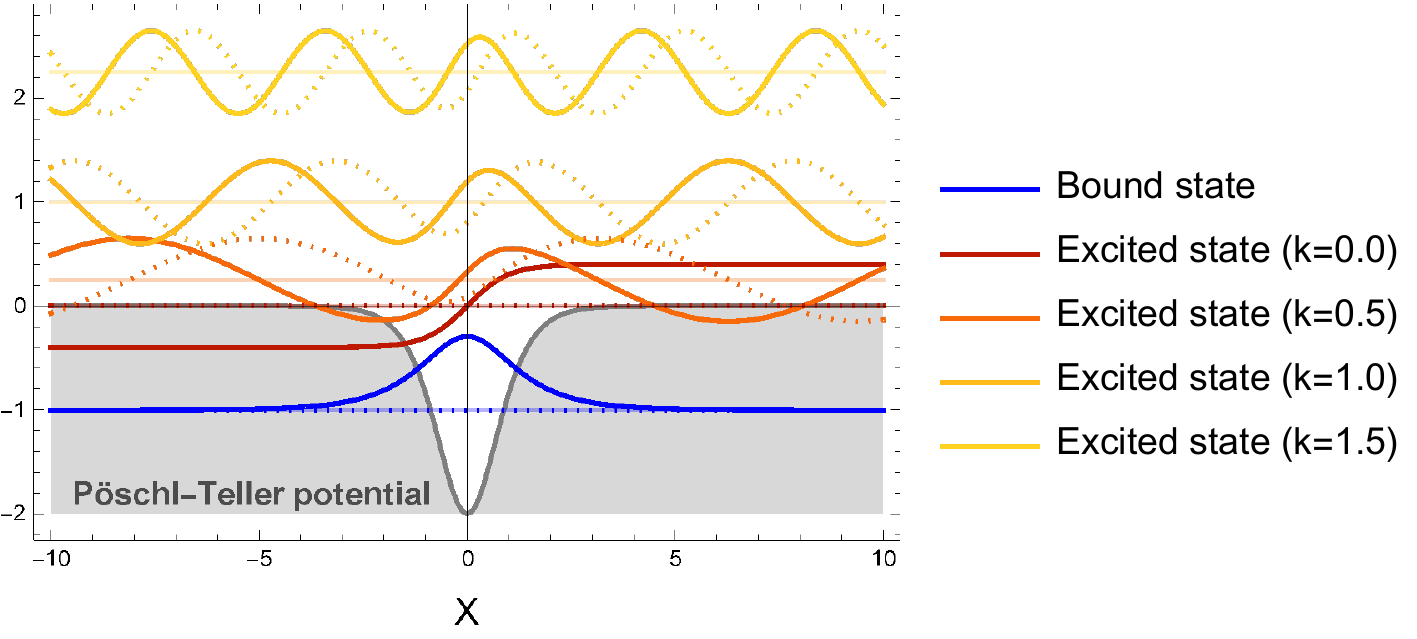}
    \caption{The eigensystem of the particle in the \PT potential. The gray line shows the form of the \PT potential. Thick blue lines show the bound state wavefunction; the solid line shows the real part, while the dotted line shows the imaginary part. A translucent blue line on the background shows the bound energy. Similarly, the other
    lines show the case of the excited states for various $k$.}
    \label{fig:enter-label}
\end{figure}

To obtain the bound state solution, let us consider the case of negative eigenenergy:
\begin{align}
    E=-\frac{\hbar^2}{2mL^2}\kappa^2\,,
\end{align}
where $\kappa$ is a real number.
Then the general solution of the above equation is given by
\begin{align}
    \psi(X) = c_1\,P_1^{\kappa}(\tanh(X)) + c_2\,Q_1^{\kappa}(\tanh(X))\,,
\end{align}
where $c_1,~c_2$ are arbitrary constants, and $P_\ell^m(z),~Q_\ell^m(z)$ are the associated Legendre polynomials of the first and the second kind respectively.
The bound state should satisfy the boundary condition
\begin{align}
    \psi(X\to \pm \infty) \simeq 0\, ,
\end{align}
which requires $\kappa=1$ and $c_2=0$. 
Also, the normalization of 
the bound state imposes $c_1=-1/\sqrt{2}$.
Hence, the bound energy and the bound state wavefunction are given by
\begin{align}
    E=-\frac{\hbar^2}{2mL^2}=:E_b,
    \quad
    \psi(X)
    =\frac{1}{\sqrt{2}\cosh(X)}=:\psi_b(X)\,.
\end{align}
Note that since the boundary condition requires $\kappa=1$, there is only one bound state in the potential.

Next, to obtain the excited state solution, let us consider the case of positive eigenenergy:
\begin{align}
    E=\frac{\hbar^2}{2mL^2}k^2=:E_k\,,
\end{align}
where $k$ is a real number taking an arbitrary value between $-\infty$ and $\infty$.
Then the general solution of the above equation is given by
\begin{align}
    \psi(X) = d_1 \frac{\tanh(X)-ik}{1-ik}e^{ikX} + 
    d_2 \frac{\tanh(X)+ik}{1+ik}e^{-ikX}\,,
\end{align}
where $d_1,~d_2$ are arbitrary constants.
The two terms are identical except that the sign of $k$ is flipped. Hence, excited state wavefunction is generally given by
\begin{align}
    \psi(X)=\frac{\tanh(X)-ik}{\sqrt{2\pi}(1-ik)}e^{ikX}=:\psi_k(X)\,.
\end{align}
The normalization of $\psi_k(X)$ will be described soon in the following paragraph.

Here, we show that the bound state $|b\rangle$ and the excited state $|k\rangle$ are orthogonal to each other as follows.
\begin{align}
    \langle b| k\rangle
    &=\int_{-\infty}^{\infty}dX \psi_b^*(X)\psi_k(X)\,,\\
    &=\left.\frac{-1}{2\sqrt{\pi}(1-ik)}\frac{e^{ikX}}{\cosh(X)}\right|_{-\infty}^\infty
    =\frac{-i\sqrt{\pi}}{\sqrt{\pi}(1-ik)}\lim_{X\to\infty}\frac{\sin(kX)}{\cosh(X)}\,.\\
    &\longrightarrow 0\
\end{align}
The sets of the excited states $|k\rangle$ also form an orthogonal basis and are normalized properly as follows.
\begin{align}
    \langle k'|k\rangle
    &=\int_{-\infty}^{\infty}dX \psi_{k'}^*(X)\psi_k(X)\,,\\
    &=\frac{1}{\pi} e^{i\phi}\lim_{X\to\infty}\frac{\sin[(k-k')X-\phi]}{k-k'}\,,\\
    &\longrightarrow \delta(k-k')\,.
\end{align}
Here, $\phi:=-\arg[(1-ik)(1+ik')]$. 
To summarize, the eigenstates of the \PT potential particle satisfy
\begin{align}
    \langle b|b\rangle=1,\quad
    \langle b|k\rangle =0,\quad
    \langle k'|k\rangle = \delta(k-k')\,.
\end{align}

\section{Saturation Time}
\label{sec:t_saturation}

Here we discuss in detail 
the calculation of the excitation probability given in Eq.~\eqref{eq:excitation_probability}.
The integrand consists of $|J_k|^2$ and $\Delta_k(t)$.
$|J_k|^2$ has a peak at $k=0$ with an $\mathcal{O}(1)$ width, while $\Delta_k(t)$ has two peaks at $k=\pm k_\text{res}$ whose width is $\delta k \simeq \pi/(|\omega_b|k_{\text res}t)$. 
To estimate the $k$ integral in Eq.~\eqref{eq:excitation_probability}, we use the saddle-point approximation depending on two different cases; (i) when the width of $|J_k|^2$ around its peak is sufficiently small compared to the peak widths of $\Delta_k(t)$, and (ii) when the peak width of $\Delta_k(t)$ is sufficiently small compared to the peak width of $|J_k|^2$. 
The condition for the case (i) 
is that the observation time is sufficiently short $t\ll t_\text{sat}$, where we defined a saturation time in Eq.~\eqref{t_sat_def}.
Then, we can perform the saddle-point approximation around the peak of $|J_k|^2$ at $k=0$. In contrast, the case (ii) 
applies for a longer observation time $t\gg t_\text{sat}$. 
In this case, $\Delta_k(t)$ can be approximated to the summation of the delta functions at the peaks $k=\pm k_\text{res}$:
\begin{align}
    \Delta_k(t\to\infty)
    \to \frac{\pi t}{2|\omega_b|}\delta(k^2-k_\text{res}^2)
    = \frac{\pi t}{4 k_\text{res}|\omega_b|}\left(\delta(k-k_\text{res})+\delta(k+k_\text{res})\right)\,.
\end{align}
Then, we can perform the $k$ integral for these delta functions as discussed in the main text. Finally, the excitation probability under the saddle-point approximation for the case (i) and (ii) reduces to the following form.
\begin{align}\label{eq:excitation_probability_saddlepoint}
    P_{\text{ex}}(t)\simeq 
    \begin{dcases}
        \sqrt{\frac{\pi^3}{4+\pi^2}} g^2 |\alpha|^2 |\omega_b|^2\, \Delta_0(t) & 
        (\text{case (i): }t\ll t_\text{sat})\\
        \frac{\pi}{2} g^2 |\alpha|^2 |\omega_b|\,t\,\frac{\left|J_{k_\text{res}}\right|^2}{k_\text{res}} & 
        (\text{case (ii): }t\gg t_\text{sat})
    \end{dcases}\,.
\end{align}
Moreover, if we focus on the case when the particle is excited dominantly at the resonant frequency, Eq.~\eqref{eq:resonance_excitation_condition} should be satisfied. Then, the particle's excitation probability is further simplified as follows.
\begin{align}\label{eq:excitation_probability_saddlepoint2}
    P_{\text{ex}}(t)\simeq 
    \begin{dcases}
        \frac{1}{4}\sqrt{\frac{\pi^3}{4+\pi^2}} g^2 |\alpha|^2 |\omega_b|^2\, t^2 & 
        (\text{case (i): }t\ll t_\text{sat})\\
        \frac{\pi}{2} g^2 |\alpha|^2 |\omega_b|\,k_\text{res}^{-1}\,t & 
        (\text{case (ii): }t\gg t_\text{sat})
    \end{dcases}\,.
\end{align}
For the case (i), we supposed $\omega_{k_\text{res}}\,t\ll1$ and perform the Taylor expansion of $\Delta_0(t)$. For the case (ii), we used $|J_{k_\text{res}}|^2\simeq 1$. 
According to the approximated form given above, the excitation probability grows quadratic in the observation time $t$ at early time. This indicates that it is better to sustain the quantum coherence of the setup as long as possible to observe the gravity-induced excitation of the particle with high probability at early time. On the other hand, at late time, the excitation probability behaves linearly in time which means that the particle is excited at a stationary rate.
This fact indicates that we do not have to sustain the quantum coherence of our setup for a longer time than $t_\text{sat}$ to see the gravity-induced excitation of the particle; instead of maintaining the quantum system for a long time and making a single observation of the particle, we can make multiple experimental runs, each of which is relatively short in duration. In this sense, our proposal can mitigate the severe limits imposed by environmental decoherence.

\begin{figure}[htbp]
    \centering
    \includegraphics[width=0.9\linewidth]{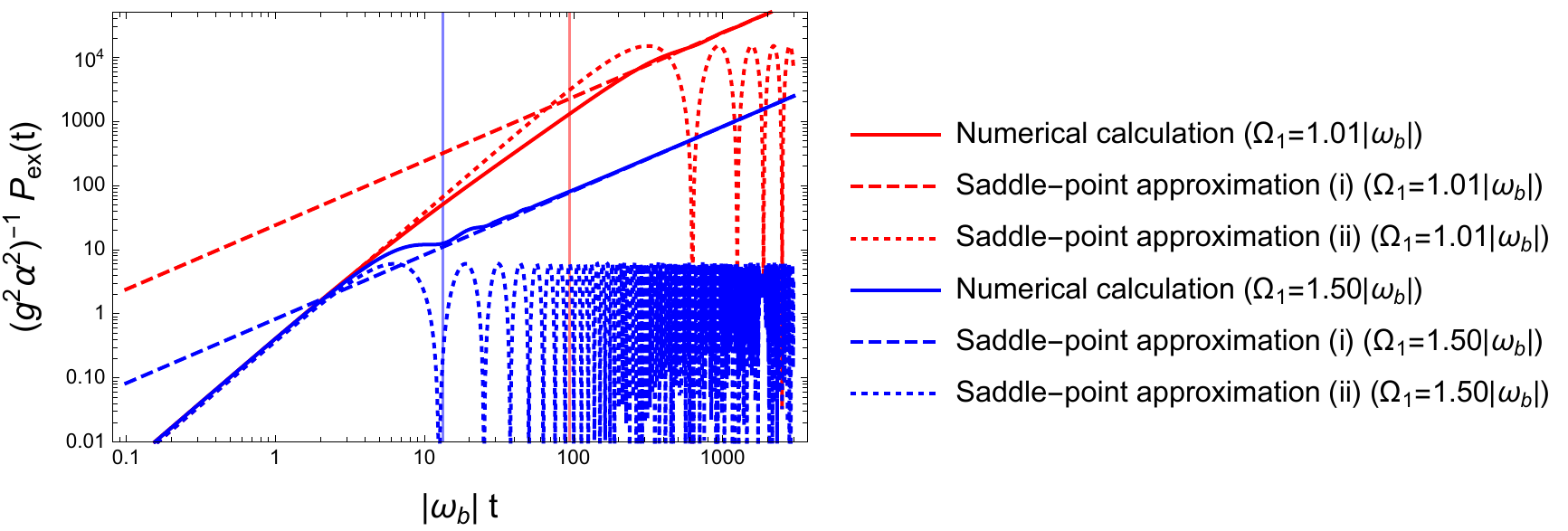}
    \caption{
    The observation time dependence of the particle's excitation probability $P_\text{ex}$. The red and blue lines show the result for $\Omega_1=1.01|\omega_b|$ and $\Omega_1=1.50|\omega_b|$, respectively. Solid lines are the numerical results, while the dashed lines and dotted lines are the analytical form obtained from the saddle-point approximation as the case (i) and (ii) in Eq.~\eqref{eq:excitation_probability_saddlepoint}, respectively. The vertical lines correspond to the saturation time scale $t_\text{sat}$ for each $\Omega_1$.
    }
    \label{fig:Pex}
\end{figure}

In Fig.~\ref{fig:Pex}, we show the time dependence of the excitation probability of the particle. The vertical axis shows the normalized excitation probability of the particle, while the horizontal axis shows the dimensionless time $|\omega_b|t$. Red and blue lines show the result for $\Omega_1=1.01|\omega_b|$ and $\Omega_1=1.50|\omega_b|$ respectively. Solid lines are the result obtained from a numerical integral of $k$ in Eq.~\eqref{eq:excitation_probability}, and the dotted and dashed lines show the saddle-point approximated result for case (i) and (ii) in Eq.~\eqref{eq:excitation_probability_saddlepoint} respectively. The saddle-point approximated results for cases (i) and (ii) agree well with the numerical results on the left and the right side of the vertical grid lines respectively, which corresponds to the saturation time $t_\text{sat}$ for each $\Omega_1$. 
Also, we can see that the excitation probability of $\Omega_1=1.01|\omega_b|$ is much greater than $\Omega_1=1.50|\omega_b|$ case, which agrees with the discussion in Eq.~\eqref{eq:resonance_excitation_condition}.


Based on the above results~\eqref{eq:excitation_probability_saddlepoint2}, let us consider the optimal duration of a single experimental run that maximizes the particle excitation probability while fixing the total experimental time. When our total experimental time is $t_{\rm tot}$ and the number of the experimental run is $N$, the duration of each run is $\tau_1 = t_{\rm tot}/N$. 
Given that the excitation probability is typically tiny, $P_{\rm ex}(\tau_1)\ll1$, for a single run, 
the probability of observing at least one excitation over the total duration is given by $P_\text{tot}:=1-(1-P_{\rm ex}(\tau_1)^N)\sim NP_{\rm ex}(\tau_1)$
. 
In the case (i), the total probability 
$P_\text{tot}\propto N \tau_1^2\propto t_{\rm tot}^2/N$

is inversely proportional to $N$. Thus it is better to decrease $N$ by extending $\tau_1$, namely, one should make the duration of a single experimental run as long as possible. Since the condition for the case (i) is $\tau_1\lesssim t_{\rm sat}$, it is preferable to have $\tau_1\simeq t_{\rm sat}$.
In the case (ii), however, the total probability 
$P_\text{tot}\propto N \tau_1=t_{\rm tot}$

depends on neither of $\tau_1$ nor $N$ as long as the total time $t_{\rm tot}$ is fixed. Then, it is experimentally favorable to shorten $\tau_1$, because maintaining quantum coherence over an extended period is generally very challenging.
Since the condition for the case (ii) is $\tau_1\gtrsim t_{\rm sat}$, it is preferable to have $\tau_1\simeq t_{\rm sat}$.
Therefore, examining both cases, we concluded that the optimal choice is $\tau_1\simeq t_{\rm sat}$.

\section{Time evolution of the particle and the qubit-oscillator system under gravitational interaction}
\label{sec:apdx_oscillator}

In this section, we derive the time-evolved state of the particle and the qubit-oscillator system given in Eq.~\eqref{eq:time_evolved_state}.
Using the Hamiltonian in Eq.~\eqref{eq:Hamiltonian}, the time-dependent \schrodinger~equation of the total system is given by
\begin{align}\label{eq:schrodinger_equation}
    i\hbar\frac{d}{d t}|\Psi(t)\rangle
    =\hat H |\Psi(t)\rangle
    =\left[\frac{\hbar\hat \Omega}{2}\left(\hat P_y^2+\hat Y^2\right) 
    + \frac{\hbar^2}{2mL^2}\left(\hat P_x^2-\frac{2}{\cosh^2(\hat X)}+g \hat X \hat Y\right)\right]|\Psi(t)\rangle\,.
\end{align}
At early times, specifically for $t<0$, the total state is given by Eq.~\eqref{eq:initial_state} as
\begin{align}\label{eq:initial_state_apdx}
    |\Psi(t<0)\rangle
    =|\Psi_b(t)\rangle 
    :=e^{-i\omega_b t}|b\rangle_{\rm p} \otimes
    \frac{1}{\sqrt{2}}\left(|0\rangle_{\rm q} |\alpha e^{-i\Omega_0 t}\rangle_{\rm o} +|1\rangle_{\rm q} |\alpha e^{-i\Omega_1 t}\rangle_{\rm o}\right)
    \,.
\end{align}
The eigendecomposition of the total state is as follows:
\begin{align}\label{eq:eigen_decomposition}
    |\Psi(t)\rangle
    &=\frac{1}{\sqrt{2}}\left(
    |0\rangle_q |\Psi_0(t)\rangle_{\rm po} + |1\rangle_q |\Psi_1(t)\rangle_{\rm po}
    \right),\\
    |\Psi_{0,1}(t)\rangle_{\rm po}
    &=\sum_{n=0}^\infty \int dj\,C_{jn}^{0,1}(t)e^{-i(\omega_j+\Omega_{0,1} n)t}|j\rangle_{\rm p} |n\rangle_{\rm o}
    \,,
\end{align}
where $C_{jn}^{0,1}$ is the decomposition coefficient depending on $\Omega_{0,1}$, $|j\rangle_p~(j=b,~k)$ represents the orthogonal eigenstate of the particle, and 
\begin{align}
    \int dj \cdots = \left(\delta_{jb} + \int_{-\infty}^\infty dk\right) \cdots 
\end{align}
denotes summation over all energy levels.
The oscillator states $|n\rangle_{\rm o}$ form the Fock basis, whose specific form depends on the eigenfrequency $\Omega_{0,1}$.
For simplicity, we omit the subscript $0,1$ of $\Omega$ and $C_{jn}$ for the qubit state labels in the following calculations.

Substituting Eq.~\eqref{eq:eigen_decomposition} into Eq.~\eqref{eq:schrodinger_equation}, the time evolution equation for the coefficients $C_{jn}(t)$ becomes:
\begin{align}\label{eq:coefficient_eom}
    \frac{d}{dt} C_{jn}(t)
    = -\frac{i\hbar^2}{2mL^2}g\,\Theta(t)\,e^{i(\omega_k+\Omega n)t}
    \sum_{n'=0}^\infty \int dj' C_{j'n'}(t) e^{-i(\omega_{k'}+\Omega n')t} \langle j|\hat X|j'\rangle_{\rm p}\,\langle n|\hat Y|n'\rangle_{\rm o}
    \,.
\end{align}
From Eq.~\eqref{eq:initial_state_apdx}, the coefficient at early time are
\begin{align}\label{eq:coefficient_earlytime}
    C_{jn}(t<0)
    =\delta_{jb}\,e^{-\alpha^2/2}\frac{\alpha^n}{\sqrt{n!}}
    \,.
\end{align}
Assuming $g$ is a small perturbation, we solve Eq.~\eqref{eq:coefficient_eom} iteratively. Substituting Eq.~\eqref{eq:coefficient_earlytime} into Eq.~\eqref{eq:coefficient_eom}, we obtain 
\begin{align}\label{eq:coef_eom}
    \frac{d}{dt} C_{jn}(t)\simeq
    -\frac{i\hbar^2}{2mL^2}g\,\Theta(t)\,e^{-\alpha^2/2}e^{i(\omega_j+\Omega n)t}
    \sum_{n'=0}^\infty\frac{\alpha^{n'}}{\sqrt{n'!}}e^{-i(\omega_b+\Omega n')t}\langle j|\hat X|b\rangle \langle n|\hat Y|n'\rangle
    +\mathcal{O}(g^2)
    \,.
\end{align}
Through straightforward calculations, we find
\begin{align}
    \langle j|\hat X|b\rangle
    &=
    \delta_{jk}\frac{\sqrt{\pi}}{2}\frac{1}{1+ik}\frac{1}{\cosh(k\pi/2)}
    =:\delta_{jk}J_k,
    \label{eq:Jk_apdx}
    \\
    \langle n|\hat Y|n'\rangle
    &=\frac{1}{\sqrt{2}}\left( \delta_{n,n'-1}\sqrt{n'}+\delta_{n,n'+1}\sqrt{n'+1}\right)
    \,.
\end{align}
By substituting these results into Eq.~\eqref{eq:coef_eom} and integrating over time $t$, we obtain
\begin{align}\label{eq:coef_sol}
    C_{jn}(t)
    =e^{-\alpha^2/2}\frac{\alpha^n}{\sqrt{n!}}\left\{
    \delta_{jb}+
    \frac{|\omega_b|}{\sqrt{2}}
    g \,\delta_{jk}J_k\left(
    \frac{n}{\alpha}
    \frac{1-e^{i(\omega_k-\omega_b+\Omega)t}}{\omega_k-\omega_b+\Omega}
    +\alpha
    \frac{1-e^{i(\omega_k-\omega_b-\Omega)t}}{\omega_k-\omega_b-\Omega}
    \right)\right\}
\end{align}
The first term of this solution yields from an integration constant which is determined by the initial condition~\eqref{eq:initial_state_apdx}.

Finally, substituting Eq.~\eqref{eq:coef_sol} into Eq.~\eqref{eq:eigen_decomposition}, we obtain the explicit form of the time-evolved state as follows.
\begin{align}\label{eq:time_evolved_state_apdx}
    |\Psi(t)\rangle
    &= |\Psi_b(t)\rangle
    + g\,\alpha\,
    |\text{ex}(\Omega_1)\rangle_{\rm p} \otimes \frac{1}{\sqrt{2}}|1\rangle_{\rm q} |\alpha e^{-i\Omega_1 t}\rangle_{\rm o}
    +g\, |\text{off-res}\rangle
    +\mathcal{O}(g^2)\,,\\
    |\text{off-res}\rangle
    &=
    \frac{1}{\sqrt{2}}\left(
    \alpha\,
    |\text{ex}(\Omega_0)\rangle_{\rm p} |0\rangle_{\rm q} |\alpha e^{-i\Omega_0 t}\rangle_{\rm o}\right.\notag\\
    &\hspace{20mm}\left.
    +
    e^{-i\Omega_0 t}\,
    |\mathrm{ex}(-\Omega_0)\rangle_{\rm p} |0\rangle_{\rm q}\, \hat a^\dagger|\alpha e^{-i\Omega_0 t}\rangle_{\rm o}
    +e^{-i\Omega_1 t}\,|\mathrm{ex}(-\Omega_1)\rangle_{\rm p} |1\rangle_{\rm q}\, \hat a^\dagger|\alpha e^{-i\Omega_1 t}\rangle_{\rm o}
    \right)\,,
\end{align}
where
\begin{align}
    |\mathrm{ex}(\Omega)\rangle
    :=\int_{-\infty}^{\infty} dk e^{-i\omega_k t}c_k(\Omega)|k\rangle\,,
    \qquad
    c_k(\Omega):=\frac{|\omega_b|J_k}{\sqrt{2}}\,\frac{1-e^{i(\omega_k-\omega_b-\Omega)t}}{\omega_k-\omega_b-\Omega}\,,
\end{align}
and $J_k$ is defined in Eq.~\eqref{eq:Jk_apdx}.
To derive this expression, the coherent state in the Fock basis was used
\begin{align}
    |\alpha\rangle=
    e^{-\alpha^2/2}\sum_n\frac{1}{\sqrt{n!}}\alpha^n|n\rangle\, .
\end{align}

Note that the second term in Eq.~\eqref{eq:time_evolved_state_apdx} includes the factor $\omega_k-\omega_b-\Omega_1$ in the denominator of $c_k(\Omega_1)$, which diverges at $k=k_\text{res}$. This indicates that the particle state is resonantly excited due to the gravitational interaction with the qubit-oscillator system.

In contrast, the terms in $|\text{off-res}\rangle$ represent off-resonance contributions. Specifically, the first term in $|\text{off-res}\rangle$ involves the inverse of $\omega_k-\omega_b-\Omega_0$, which remains regular under the condition in Eq.~\eqref{eq:resonance_condition}. 

To compare the contributions of the first term in $|\text{off-res}\rangle$ and the resonant term, we evaluate the ratio of their norms:
\begin{align}
    \frac{\langle \text{ex}(\Omega_0)|\text{ex}(\Omega_0)\rangle}{\langle \text{ex}(\Omega_1)|\text{ex}(\Omega_1)\rangle}
    =\frac{\int^\infty_{-\infty}dk |J_k|^2\Delta_k(\Omega_0)}{\int^\infty_{-\infty}dk |J_k|^2\Delta_k(\Omega_1)}.
    \label{eq:ratio}
\end{align}
We will derive the upper bound of this ratio by overestimating the numerator. We then require that the upper bound remains well below unity.
First, the overestimation of $\langle \text{ex}(\Omega_0)|\text{ex}(\Omega_0)\rangle$ is given by
    \begin{align}
        \langle \text{ex}(\Omega_0)|\text{ex}(\Omega_0)\rangle
        \leq 2|\omega_b|^2\Delta_0(\Omega_0) \int^{\infty}_{-\infty}dk |J_k|^2
        \simeq \sqrt{\frac{4\pi^3}{4+\pi^2}}|\omega_b|^2
        \left(\frac{\sin[(|\omega_b|-\Omega_0)t/2]}{|\omega_b|-\Omega_0}\right)^2
        ~.
    \end{align}
    Here, we used the fact that $\Delta_k(\Omega_0)$ attains its maximum at $k=0$. Also, we performed the $k$-integral of $|J_k|^2$ using a saddle point approximation at $k=0$.
    Next, $\langle \text{ex}(\Omega_1)|\text{ex}(\Omega_1)\rangle$ is approximately given by
    \begin{align}
        \langle \text{ex}(\Omega_1)|\text{ex}(\Omega_1)\rangle
        \simeq
        \frac{\pi^2}{4}\frac{|\omega_b|}{k_{\rm res}}t
        ~,
    \end{align}
    where we have assumed $t\gtrsim t_{\rm sat}=\frac{\pi}{k_{\rm res}|\omega_b|}$ and employed the saddle point approximation, as discussed in Appendix.~\ref{sec:t_saturation}.
    Thus, the upper bound of the ratio in Eq.~\eqref{eq:ratio} becomes
    \begin{align}
        \frac{\langle \text{ex}(\Omega_0)|\text{ex}(\Omega_0)\rangle}{\langle \text{ex}(\Omega_1)|\text{ex}(\Omega_1)\rangle}
        \lesssim
        \frac{|\omega_b|(\Omega_1-|\omega_b|)}{(|\omega_b|-\Omega_0)^2}
        ~.
    \end{align}
    Here, we used Eq.~\eqref{eq:resonant wavenumber} and neglected the constant factor and the sine function. Also, we again used $t\gtrsim t_{\rm sat}$.
    Hence, the sufficient condition for the ratio in Eq.~\eqref{eq:ratio} to be small is given by:
    \begin{align}
        t\gtrsim t_{\rm sat}
    \end{align}
    and
    \begin{align}
        \frac{|\omega_b|(\Omega_1-|\omega_b|)}{(|\omega_b|-\Omega_0)^2} \ll 1
        ~.
    \end{align}
The first condition implies that the observation time $t$ must be long enough to resolve the resonant peak. The second condition indicates that $\Omega_0$ should be sufficiently detuned from $|\omega_b|$, so as to suppress any competing excitation and allow exclusive resonance at $\Omega_1$.
Especially, when the difference between $\Omega_0$ and $\Omega_1$ is not very large, as in typical optomechanical setups, the second condition can be simplified as
\begin{align}
    1-\frac{|\omega_b|}{\Omega_1}\ll \left(1-\frac{\Omega_0}{\Omega_1}\right)^2~.
    \label{eq:exclusive_resonance_condition}
\end{align}
Here, we have used $\Omega_0<|\omega_b|<\Omega_1$ and $1-\frac{\Omega_0}{\Omega_1}\ll1$. This implies that the particle eigen frequency $|\omega_b|$ must be extremely close to $\Omega_1$.
Further discussion on the exclusive resonance can be found in \cite{Kaku2023}.
In the main text, we implicitly assume that the both conditions are satisfied.

The second and third terms of $|\text{off-res}\rangle$ contain the inverse of $\omega_k-\omega_b+\Omega_{0,1}$, which are also regular. 
By neglecting the off-resonance terms and higher-order corrections of $g$, the simplified expression of the time-evolved state is obtained as in Eq.~\eqref{eq:time_evolved_state_apdx}.

\section{Superposition of two coherent states in the optomechanical system}
\label{sec:apdx_optomecha}


In this section, we demonstrate that the state described in Eq.~\eqref{eq:initial_state_opm} can be prepared using the optomechanical device. The optomechanical setup is illustrated in Fig.~\ref{fig:setup2} and its Hamiltonian is given in Eq.~\eqref{eq:Hamiltonian_opm}. The oscillator system in the optomechanical device has two distinct eigenfrequencies, $\Omega_0$ and $\Omega_1:=\sqrt{\Omega_0^2+\frac{\hbar\omega_c}{M \ell^2}}$.

At the initial time $t_{\text{ini}}(<0)$, we prepare the state
\begin{align}
    |\psi(t_{\text{ini}})\rangle_{\text{opm}}
    =\frac{1}{\sqrt{2}}\left(|0\rangle+|1\rangle\right)\otimes |\alpha\rangle
    \,,
\end{align}
where $|\alpha\rangle$ is an initial coherent state of the oscillator with frequency $\Omega_0$.
As the system evolves, the state at time $t$ becomes
\begin{align}
    |\psi(t)\rangle
    =\frac{1}{\sqrt{2}}e^{-i\omega_c (t-t_{\rm ini})}\left(
    e^{i\phi_0}|0\rangle|\alpha_0,\zeta_0\rangle
    +e^{i\phi_1}|1\rangle|\alpha_1,\zeta_1\rangle
    \right)
    \, ,
\end{align}
where $|\alpha_{n_1},\zeta_{n_1}\rangle$ represents a coherent squeezed state, and $\phi_{n_1}$ is the phase factor. Their explicit forms are
\begin{align}
    \alpha_{n_1}(t)
    &=
    \begin{dcases}
        ~ e^{-i\Omega_0 (t-t_{\rm ini})}\alpha & \hspace{15mm} (n_1=0)\\
        ~ \sqrt{\frac{\Omega_1}{\Omega_0}}\left\{
        e^{-i\Omega_1 (t-t_{\rm ini})}\alpha + (1-e^{-i\Omega_1 (t-t_{\rm ini})})
        \left(\frac{\Omega_0}{\Omega_1}\right)^2\lambda_0
        \right\}& \hspace{15mm} (n_1=1)
    \end{dcases}
    \quad ,\\
    \zeta_{n_1}(t)
    &=
    \begin{dcases}
        ~ 0 & \hspace{56mm} (n_1=0)\\
        ~ e^{-2i\Omega_1 (t-t_{\rm ini})}\log\left[\sqrt{\Omega_1/\Omega_0}\right] & \hspace{56mm} (n_1=1)
    \end{dcases}
    \quad ,\\
    \phi_{n_1}(t)
    &=
    \begin{dcases}
        ~ -\frac{1}{2}\Omega_0 (t-t_{\rm ini}) &  (n_1=0)\\
        ~ -\frac{1}{2}\Omega_1 (t-t_{\rm ini})
        + \left(\frac{\Omega_0}{\Omega_1}\right)^{3/2} \lambda_0 \,\mathrm{Im}[\alpha_1(1-e^{-i\Omega_1 (t-t_{\rm ini})})] + \mathcal{O}(\lambda_0^2)
        &  (n_1=1)
    \end{dcases}
    \quad .
\end{align}
Here, $\lambda_0:=\frac{\omega_c}{\Omega_0}\frac{1}{\ell}\sqrt{\frac{\hbar}{2M\Omega_0}}$ is the optomechanical coupling constant.

Next, let us consider the limit where the optomechanical interaction occurs over a sufficiently long duration, i.e. $t-t_{\text{ini}}\to \infty$. Additionally, we assume that the optomechanical interaction is weak and take the limit of $\Omega_1\to \Omega_0$, while keeping $(\Omega_1-\Omega_0)(t-t_{\text{ini}})$ constant. Under these conditions, the time-evolved state simplifies to
\begin{align}\label{eq:initial_state_opm_apdx}
    |\psi(t)\rangle_{\text{opm}}
    \simeq \frac{1}{\sqrt{2}}e^{-i\omega_c (t-t_{\rm ini})}\left(
    e^{-i\Omega_0 (t-t_{\rm ini})/2}
    |0\rangle|\alpha_0(t)\rangle
    +e^{-i\Omega_1 (t-t_{\rm ini})/2}
    |1\rangle|\alpha_1(t)\rangle\right)
    \, ,
\end{align}
where
\begin{align}
    \alpha_{n_1}(t)
    \sim
    \begin{dcases}
        e^{-i\Omega_0 (t-t_{\rm ini})}\alpha & (n_1=0)\\
        e^{-i\Omega_1 (t-t_{\rm ini})}(\alpha-\lambda_0) + \lambda_0 & (n_1=1)
    \end{dcases}
    \quad .
\end{align}
Thus, the optomechanical setup allows us to prepare a superposition of two coherent states with different eigenfrequencies, $\Omega_0$ and $\Omega_1$. In Eq.~\eqref{eq:initial_state_opm} of the main text, the phase factors in Eq.~\eqref{eq:initial_state_opm_apdx} were omitted for simplicity, as they do not affect the primary results.

\bibliographystyle{unsrt}
\bibliography{reference}

\begin{thebibliography}{10}

\bibitem{Feynmann}
C{\'e}cile~M DeWitt and Dean Rickles.
\newblock {\em The role of gravitation in physics: report from the 1957 Chapel Hill Conference}.
\newblock Edition Open Access, 2011.

\bibitem{Bose2017}
Sougato Bose, Anupam Mazumdar, Gavin~W Morley, Hendrik Ulbricht, Marko Toro{\v{s}}, Mauro Paternostro, Andrew~A Geraci, Peter~F Barker, MS~Kim, and Gerard Milburn.
\newblock Spin entanglement witness for quantum gravity.
\newblock {\em Physical review letters}, 119(24):240401, 2017.

\bibitem{Marletto2017}
Chiara Marletto and Vlatko Vedral.
\newblock Gravitationally induced entanglement between two massive particles is sufficient evidence of quantum effects in gravity.
\newblock {\em Physical review letters}, 119(24):240402, 2017.

\bibitem{Kibble1978}
TWB Kibble.
\newblock Relativistic models of nonlinear quantum mechanics.
\newblock {\em Communications in Mathematical Physics}, 64:73--82, 1978.

\bibitem{Kibble1980}
TWB Kibble and S~Randjbar-Daemi.
\newblock Non-linear coupling of quantum theory and classical gravity.
\newblock {\em Journal of Physics A: Mathematical and General}, 13(1):141, 1980.

\bibitem{Diosi1989}
Lajos Di{\'o}si.
\newblock Models for universal reduction of macroscopic quantum fluctuations.
\newblock {\em Physical Review A}, 40(3):1165, 1989.

\bibitem{Diosi2011}
Lajos Di{\'o}si.
\newblock The gravity-related decoherence master equation from hybrid dynamics.
\newblock In {\em Journal of Physics: Conference Series}, volume 306, page 012006. IOP Publishing, 2011.

\bibitem{Penrose1996}
Roger Penrose.
\newblock On gravity's role in quantum state reduction.
\newblock {\em General relativity and gravitation}, 28:581--600, 1996.

\bibitem{Penrose2014}
Roger Penrose.
\newblock On the gravitization of quantum mechanics 1: Quantum state reduction.
\newblock {\em Foundations of Physics}, 44:557--575, 2014.

\bibitem{Kafri2014}
D~Kafri, JM~Taylor, and GJ~Milburn.
\newblock A classical channel model for gravitational decoherence.
\newblock {\em New Journal of Physics}, 16(6):065020, 2014.

\bibitem{Tilloy2016}
Antoine Tilloy and Lajos Di{\'o}si.
\newblock Sourcing semiclassical gravity from spontaneously localized quantum matter.
\newblock {\em Physical Review D}, 93(2):024026, 2016.

\bibitem{Bassi2017}
Angelo Bassi, Andr{\'e} Gro{\ss}ardt, and Hendrik Ulbricht.
\newblock Gravitational decoherence.
\newblock {\em Classical and Quantum Gravity}, 34(19):193002, 2017.

\bibitem{Carney2023}
Daniel Carney and Jacob~M Taylor.
\newblock Strongly incoherent gravity.
\newblock {\em arXiv preprint arXiv:2301.08378}, 2023.

\bibitem{Oppenheim2023}
Jonathan Oppenheim.
\newblock A postquantum theory of classical gravity?
\newblock {\em Physical Review X}, 13(4):041040, 2023.

\bibitem{Bahrami2014}
Mohammad Bahrami, Andr{\'e} Gro{\ss}ardt, Sandro Donadi, and Angelo Bassi.
\newblock The schr{\"o}dinger--newton equation and its foundations.
\newblock {\em New Journal of Physics}, 16(11):115007, 2014.

\bibitem{Anastopoulos2014}
C~Anastopoulos and BL~Hu.
\newblock Problems with the newton--schr{\"o}dinger equations.
\newblock {\em New Journal of Physics}, 16(8):085007, 2014.

\bibitem{Ruffini1969}
Remo Ruffini and Silvano Bonazzola.
\newblock Systems of self-gravitating particles in general relativity and the concept of an equation of state.
\newblock {\em Phys. Rev.}, 187:1767--1783, Nov 1969.

\bibitem{Rijavec2021}
Simone Rijavec, Matteo Carlesso, Angelo Bassi, Vlatko Vedral, and Chiara Marletto.
\newblock Decoherence effects in non-classicality tests of gravity.
\newblock {\em New Journal of Physics}, 23(4):043040, 2021.

\bibitem{Panda2024}
Cristian~D Panda, Matthew Tao, James Egelhoff, Miguel Ceja, Victoria Xu, and Holger M{\"u}ller.
\newblock Coherence limits in lattice atom interferometry at the one-minute scale.
\newblock {\em Nature Physics}, pages 1--6, 2024.

\bibitem{Bild2023}
Marius Bild, Matteo Fadel, Yu~Yang, Uwe Von~L{\"u}pke, Phillip Martin, Alessandro Bruno, and Yiwen Chu.
\newblock Schr{\"o}dinger cat states of a 16-microgram mechanical oscillator.
\newblock {\em Science}, 380(6642):274--278, 2023.

\bibitem{Fein2019}
Yaakov~Y Fein, Philipp Geyer, Patrick Zwick, Filip Kia{\l}ka, Sebastian Pedalino, Marcel Mayor, Stefan Gerlich, and Markus Arndt.
\newblock Quantum superposition of molecules beyond 25 kda.
\newblock {\em Nature Physics}, 15(12):1242--1245, 2019.

\bibitem{Westphal2021}
Tobias Westphal, Hans Hepach, Jeremias Pfaff, and Markus Aspelmeyer.
\newblock Measurement of gravitational coupling between millimetre-sized masses.
\newblock {\em Nature}, 591(7849):225--228, 2021.

\bibitem{Lee2020}
JG~Lee, EG~Adelberger, TS~Cook, SM~Fleischer, and BR~Heckel.
\newblock New test of the gravitational 1/r 2 law at separations down to 52 $\mu$ m.
\newblock {\em Physical Review Letters}, 124(10):101101, 2020.

\bibitem{Krisnanda2020}
Tanjung Krisnanda, Guo~Yao Tham, Mauro Paternostro, and Tomasz Paterek.
\newblock Observable quantum entanglement due to gravity.
\newblock {\em npj Quantum Information}, 6(1):12, 2020.

\bibitem{Fujita2023}
Tomohiro Fujita, Youka Kaku, Akira Matsumura, and Yuta Michimura.
\newblock Inverted oscillators for testing gravity-induced quantum entanglement.
\newblock {\em arXiv preprint arXiv:2308.14552}, 2023.

\bibitem{Kaku2023}
Youka Kaku, Tomohiro Fujita, and Akira Matsumura.
\newblock Enhancement of quantum gravity signal in an optomechanical experiment.
\newblock {\em Physical Review D}, 108(10):106014, 2023.

\bibitem{Pedernales2022}
Julen~S. Pedernales, Kirill Streltsov, and Martin~B. Plenio.
\newblock Enhancing gravitational interaction between quantum systems by a massive mediator.
\newblock {\em Phys. Rev. Lett.}, 128:110401, Mar 2022.

\bibitem{Poschl1933}
Herta P\"oschl and Edward Teller.
\newblock Bemerkungen zur quantenmechanik des anharmonischen oszillators.
\newblock {\em Z. Physik}, 83:143--151, 1933.

\bibitem{Zyczkowski1998}
Karol {\.Z}yczkowski, Pawe{\l} Horodecki, Anna Sanpera, and Maciej Lewenstein.
\newblock Volume of the set of separable states.
\newblock {\em Physical Review A}, 58(2):883, 1998.

\bibitem{Eisert2003}
Jens Eisert and MB1069 Plenio.
\newblock Introduction to the basics of entanglement theory in continuous-variable systems.
\newblock {\em International Journal of Quantum Information}, 1(04):479--506, 2003.

\bibitem{Aspelmeyer2014}
Markus Aspelmeyer, Tobias~J. Kippenberg, and Florian Marquardt.
\newblock Cavity optomechanics.
\newblock {\em Rev. Mod. Phys.}, 86:1391--1452, Dec 2014.

\bibitem{Janitz2015}
Erika Janitz, Maximilian Ruf, Mark Dimock, Alexandre Bourassa, Jack Sankey, and Lilian Childress.
\newblock Fabry-perot microcavity for diamond-based photonics.
\newblock {\em Phys. Rev. A}, 92:043844, Oct 2015.

\end{thebibliography}


\begin{thebibliography}{10}

\bibitem{Poschl:1933zz}
G.~Poschl and E.~Teller,
``Bemerkungen zur Quantenmechanik des anharmonischen Oszillators,''
Z. Phys. \textbf{83}, 143-151 (1933)
doi:10.1007/BF01331132

\bibitem{Flügge:1999}
Siegfried Flügge,
``Practical Quantum Mechanics,''
Springer Berlin, Heidelberg (1933)
doi:10.1007/978-3-642-61995-3

\bibitem{Panda2024}
Panda, C.D., Tao, M., Egelhoff, J. et al.,
``Coherence limits in lattice atom interferometry at the one-minute scale'',
Nat. Phys. (2024),
doi:10.1038/s41567-024-02518-9


\end{thebibliography}

\end{document}